\documentstyle[prl,eqsecnum,psfig,aps]{revtex}
\input{epsf}       
\begin{document}

\begin{flushright}
{\bf Fermilab-Pub-99/139-E} \\
{\bf D0Pub-99-1}
\end{flushright}
\vskip 0.5cm
\begin{center}
\Large \bf Studies of $WW$ and $WZ$ Production and Limits on Anomalous
$WW\gamma$ and $WWZ$ Couplings
\end{center}
\vskip 1.0cm
\centerline{The D\O \ Collaboration$^*$}
\centerline{\it Fermi National Accelerator Laboratory, Batavia, IL 60510}
\centerline{(May 4, 1999)}
\vskip 1.5cm
\centerline{\Large Abstract}
\vskip 1.0cm
Evidence of anomalous $WW$ and $WZ$ production was sought in
$p \bar{p}$ collisions at a center-of-mass energy of $\sqrt{s} = 1.8$ TeV.
The final states $WW (WZ)\rightarrow \mu\nu
\;{\rm jet \; jet}+X$, $WZ\rightarrow \mu\nu ee+X$ and
$WZ\rightarrow e\nu ee+X$ 
were studied using a data sample corresponding to an
integrated luminosity of approximately $90~{\rm pb}^{-1}$.
No evidence of anomalous diboson production was found.  Limits were set on
anomalous $WW\gamma$ and $WWZ$ couplings and were combined with our previous
results. The combined $95\%$ confidence level 
anomalous coupling limits for $\Lambda=2$
TeV are $-0.25 \le \Delta \kappa \le 0.39 \; (\lambda=0)$ and
$-0.18 \le \lambda \le 0.19 \; (\Delta \kappa = 0)$, assuming the
$WW\gamma$ couplings are equal to the $WWZ$ couplings.

\begin{flushleft}
------ \\
$^*$ Authors listed on the following page. \\
Submitted to Physical Review D.
\end{flushleft}
\global\advance\count0 -1
\clearpage

\title{Studies of $WW$ and $WZ$ Production and Limits on Anomalous 
$WW\gamma$ and $WWZ$ Couplings}
%
%
\author{                                                                      
B.~Abbott,$^{45}$                                                             
M.~Abolins,$^{42}$                                                            
V.~Abramov,$^{18}$                                                            
B.S.~Acharya,$^{11}$                                                          
I.~Adam,$^{44}$                                                               
D.L.~Adams,$^{54}$                                                            
M.~Adams,$^{28}$                                                              
S.~Ahn,$^{27}$                                                                
V.~Akimov,$^{16}$                                                             
G.A.~Alves,$^{2}$
% ...
\\                                                                            
\vskip 0.05cm                                                                 
\centerline{(D\O\ Collaboration)}                                             
\vskip 0.05cm                                                                 
}                                                                             
\address{                                                                     
\centerline{$^{1}$Universidad de Buenos Aires, Buenos Aires, Argentina}       
\centerline{$^{2}$LAFEX, Centro Brasileiro de Pesquisas F{\'\i}sicas,         
                  Rio de Janeiro, Brazil}                                     
\centerline{$^{3}$Universidade do Estado do Rio de Janeiro,                   
                  Rio de Janeiro, Brazil}                                     
\centerline{$^{4}$Institute of High Energy Physics, Beijing,                  
                  People's Republic of China}                                 
\centerline{$^{5}$Universidad de los Andes, Bogot\'{a}, Colombia}             
\centerline{$^{6}$Universidad San Francisco de Quito, Quito, Ecuador}         
\centerline{$^{7}$Institut des Sciences Nucl\'eaires, IN2P3-CNRS,             
                  Universite de Grenoble 1, Grenoble, France}                 
\centerline{$^{8}$DAPNIA/Service de Physique des Particules, CEA, Saclay,     
                  France}                                                     
\centerline{$^{9}$Panjab University, Chandigarh, India}                       
\centerline{$^{10}$Delhi University, Delhi, India}                            
\centerline{$^{11}$Tata Institute of Fundamental Research, Mumbai, India}     
\centerline{$^{12}$Kyungsung University, Pusan, Korea}                        
\centerline{$^{13}$Seoul National University, Seoul, Korea}                   
\centerline{$^{14}$CINVESTAV, Mexico City, Mexico}                            
\centerline{$^{15}$Institute of Nuclear Physics, Krak\'ow, Poland}            
\centerline{$^{16}$Institute for Theoretical and Experimental Physics,        
                   Moscow, Russia}                                            
\centerline{$^{17}$Moscow State University, Moscow, Russia}                   
\centerline{$^{18}$Institute for High Energy Physics, Protvino, Russia}       
\centerline{$^{19}$Lancaster University, Lancaster, United Kingdom}           
\centerline{$^{20}$University of Arizona, Tucson, Arizona 85721}              
\centerline{$^{21}$Lawrence Berkeley National Laboratory and University of    
                   California, Berkeley, California 94720}                    
\centerline{$^{22}$University of California, Davis, California 95616}         
\centerline{$^{23}$University of California, Irvine, California 92697}        
\centerline{$^{24}$University of California, Riverside, California 92521}     
\centerline{$^{25}$Florida State University, Tallahassee, Florida 32306}      
\centerline{$^{26}$University of Hawaii, Honolulu, Hawaii 96822}              
\centerline{$^{27}$Fermi National Accelerator Laboratory, Batavia,            
                   Illinois 60510}                                            
\centerline{$^{28}$University of Illinois at Chicago, Chicago,                
                   Illinois 60607}                                            
\centerline{$^{29}$Northern Illinois University, DeKalb, Illinois 60115}      
\centerline{$^{30}$Northwestern University, Evanston, Illinois 60208}         
\centerline{$^{31}$Indiana University, Bloomington, Indiana 47405}            
\centerline{$^{32}$University of Notre Dame, Notre Dame, Indiana 46556}       
\centerline{$^{33}$Purdue University, West Lafayette, Indiana 47907}          
\centerline{$^{34}$Iowa State University, Ames, Iowa 50011}                   
\centerline{$^{35}$University of Kansas, Lawrence, Kansas 66045}              
\centerline{$^{36}$Kansas State University, Manhattan, Kansas 66506}          
\centerline{$^{37}$Louisiana Tech University, Ruston, Louisiana 71272}        
\centerline{$^{38}$University of Maryland, College Park, Maryland 20742}      
\centerline{$^{39}$Boston University, Boston, Massachusetts 02215}            
\centerline{$^{40}$Northeastern University, Boston, Massachusetts 02115}      
\centerline{$^{41}$University of Michigan, Ann Arbor, Michigan 48109}         
\centerline{$^{42}$Michigan State University, East Lansing, Michigan 48824}   
\centerline{$^{43}$University of Nebraska, Lincoln, Nebraska 68588}           
\centerline{$^{44}$Columbia University, New York, New York 10027}             
\centerline{$^{45}$New York University, New York, New York 10003}             
\centerline{$^{46}$University of Rochester, Rochester, New York 14627}        
\centerline{$^{47}$State University of New York, Stony Brook,                 
                   New York 11794}                                            
\centerline{$^{48}$Brookhaven National Laboratory, Upton, New York 11973}     
\centerline{$^{49}$Langston University, Langston, Oklahoma 73050}             
\centerline{$^{50}$University of Oklahoma, Norman, Oklahoma 73019}            
\centerline{$^{51}$Brown University, Providence, Rhode Island 02912}          
\centerline{$^{52}$University of Texas, Arlington, Texas 76019}               
\centerline{$^{53}$Texas A\&M University, College Station, Texas 77843}       
\centerline{$^{54}$Rice University, Houston, Texas 77005}                     
}                                                                             
\maketitle
\begin{abstract}
Evidence of anomalous $WW$ and $WZ$ production was sought in
$p \bar{p}$ collisions at a center-of-mass energy of $\sqrt{s} = 1.8$ TeV.
The final states $WW (WZ)\rightarrow \mu\nu
\;{\rm jet \; jet}+X$, $WZ\rightarrow \mu\nu ee+X$ and
$WZ\rightarrow e\nu ee+X$ 
were studied using a data sample corresponding to an
integrated luminosity of approximately $90~{\rm pb}^{-1}$.
No evidence of anomalous diboson production was found.  Limits were set on
anomalous $WW\gamma$ and $WWZ$ couplings and were combined with our previous
results. The combined $95\%$ confidence level 
anomalous coupling limits for $\Lambda=2$
TeV are $-0.25 \le \Delta \kappa \le 0.39 \; (\lambda=0)$ and
$-0.18 \le \lambda \le 0.19 \; (\Delta \kappa = 0)$, assuming the
$WW\gamma$ couplings are equal to the $WWZ$ couplings.
\end{abstract}
%
%
\twocolumn
\vskip 2.0 cm
\normalsize 

\section{Introduction}
\label{sec-intro}
The gauge theory of the electroweak interactions contains a striking
feature.  Unlike the electrically neutral photon in quantum electrodynamics 
(QED), the weak vector
bosons carry weak charge.  Consequently, whereas in QED there are no
photon-photon couplings, the weak vector bosons interact amongst themselves
through the trilinear and quartic gauge boson vertices.
                                     
A formalism has been developed to describe the $WW\gamma$ and $WWZ$ vertices
for the most general gauge boson self-interactions \cite{lagrangian,HWZ}.
The Lorentz invariant effective Lagrangian for the gauge boson
self-interactions
contains fourteen dimensionless couplings, seven each for
$WW\gamma$ and $WWZ$:
\begin{eqnarray*}
 {\cal L}_{WWV}/g_{WWV} =
   ig_1^V \left( W^{\dag}_{\mu\nu}W^{\mu}V^{\nu}
  -  W^{\dag}_{\mu}V_{\nu}W^{\mu\nu} \right) \\ +
i\kappa_V W^{\dag}_{\mu}W_{\nu}
  V^{\mu\nu} + i\frac{\lambda_V}{M_W^2} W_{\lambda\mu}^{\dag}W_{\nu}^{\mu}
  V^{\nu\lambda} \nonumber \\
  -g^V_4W^{\dag}_{\mu}W_{\nu}(\partial^{\mu}V^{\nu} + \partial^{\nu}V^{\mu}) \\
  +g^V_5\epsilon^{\mu\nu\rho\alpha} \left( W^{\dag}_{\mu}
  \stackrel{\leftrightarrow}{\partial}_{\rho}
  W_{\nu} \right)V_{\alpha} \nonumber \\
  + i\tilde{\kappa}_V
   W_{\mu}^{\dag}W_{\nu}\tilde{V}^{\mu\nu}
   +\frac{i\tilde{\lambda}_V}{M_W^2}W^{\dag}_{\lambda\mu}W^{\mu}_{\nu}
   \tilde{V}^ {\nu\lambda} ,
\end{eqnarray*}
where $W^{\mu}$ denotes the $W^-$ field, $W_{\mu\nu}=\partial_{\mu}W_{\nu}-
\partial_{\nu}W_{\mu}$, $V_{\mu\nu} = \partial_{\mu}V_{\nu}-
\partial_{\nu}V_{\mu}$, $\tilde{V}_{\mu\nu}=\frac{1}{2} \epsilon_
{\mu\nu\rho\alpha}V^{\rho\alpha}$, and $(A\stackrel{\leftrightarrow}\partial
_{\mu}B)=A(\partial_{\mu}B)-(\partial_{\mu}A)B$, $V=\gamma$ and $Z$, and
$M_W$ is the mass of the $W$ boson. 
The overall coupling parameters $g_{WWV}$ are
$g_{WW\gamma} = -e$ and $g_{WWZ} = -e\, {\rm cot} \theta _{w}$, as in the 
standard model (SM),
where $e$ and $\theta_w$ are the positron charge and the weak mixing
angle. 
The couplings $\lambda_V$ and $\kappa_V$ conserve $C$ and $P$.
The couplings $g_4^V$ are odd under $CP$ and $C$, $g_5^V$ are
odd under $C$ and $P$, and $\tilde{\kappa}_V$ and $\tilde{\lambda}_V$ are odd
under $CP$ and $P$.
In the SM, all the couplings are zero at tree level with the exception of 
$g_1^V$ and $\kappa_V$ ($g_1^{\gamma} = g_1^Z = \kappa_{\gamma} = \kappa_Z 
= 1$), and $\Delta\kappa_V$ and $\Delta g_1^Z$ are defined as $\kappa_V-1$ and
$g_1^Z-1$, respectively.
Electromagnetic gauge invariance restricts
$g_1^{\gamma}, g_4^{\gamma}$, and $g_5^{\gamma}$ to the SM
values of 1, 0, and 0.
The $CP$-violating $WW\gamma$ couplings $\tilde{\lambda}_\gamma$ and
$\tilde{\kappa}_\gamma$ have been tightly constrained by measurements of the
neutron electric dipole moment to
$|\tilde{\kappa}_{\gamma}|,|\tilde{\lambda}_{\gamma}| < 10^{-3}$
\cite{nedm}.

With non-SM coupling parameters, the amplitudes for gauge boson
pair production grow with energy, eventually violating tree-level
unitarity.  The unitarity violation is avoided by
parameterizing the anomalous couplings as dipole form factors with a cutoff
scale, $\Lambda$.  Then the anomalous couplings take a form, for example,
\begin{eqnarray*}
\Delta\kappa(\hat{s}) = \frac{\Delta\kappa}{(1+ \hat{s}/\Lambda^2)^2},
\end{eqnarray*}
where $\hat{s}$ is the invariant mass of the vector boson pair and 
$\Delta \kappa$ is the coupling value at the low energy limit~\cite{newref}. 
$\Lambda$
is physically interpreted as the mass scale where the new phenomenon which
is responsible for the anomalous couplings would be directly observable. 
              
Direct tests of the trilinear couplings are provided by $e^+e^-$ and $p\bar{p}$
colliders through production of gauge boson pairs, in particular by
$e^+e^-\rightarrow W^+W^-$, $Z\gamma$, and $ZZ$ and by $p\bar{p}\rightarrow
W^{\pm}\gamma$, $W^+W^-$, $W^{\pm}Z$, $Z\gamma$, and $ZZ$.
The experiments seek to measure, or otherwise place limits on, trilinear 
couplings and to retain
sensitivity to the appearance of new physical phenomena.
The signature for anomalous trilinear couplings is an excess
of gauge boson pairs, particularly for large values of the invariant mass
of the gauge boson pair and for large values of gauge boson
transverse momentum $p_T$.

Limits on these couplings are often obtained under the assumption
that the $WW\gamma$ and $WWZ$ couplings are equal $(g_1^{\gamma}=g_1^Z,
\Delta\kappa_{\gamma}=\Delta\kappa_Z$, and $\lambda_{\gamma}=\lambda_Z)$.
Another set of parameters, $\alpha_{B\phi}$, $\alpha_{W\phi}$, and $\alpha_W$,
is similarly motivated by $SU(2)_{L} \times U(1)_{Y}$ gauge
invariance.  These couplings are linear combinations of $\lambda_V$,
$\Delta\kappa_V$, and $\Delta g_1^Z$ such that
$\alpha_{B\phi}=\Delta\kappa_{\gamma}-\Delta g_1^Z \cos^2{\theta_w}$,
$\alpha_{W\phi}=\Delta g_1^Z \cos^2{\theta_w}$, and
$\alpha_W=\lambda_{\gamma}$ with the constraints that
$\Delta\kappa_Z=-\Delta\kappa_{\gamma} \tan^2{\theta_w}+\Delta g_1^Z$
and $\lambda_{\gamma}=\lambda_Z$.  Adding the additional constraint that
$\alpha_{B\phi}=\alpha_{W\phi}$ yields~\cite{HISZ} the HISZ relations 
used by the D\O \ and CDF collaborations.

The D\O \ collaboration has previously performed several searches for
anomalous $WW\gamma$ and $WWZ$ couplings.  Studies~\cite{D0WG01,D0WG02} of
$p\bar{p}\rightarrow W\gamma +X$ have shown that the transverse energy spectrum
of the photons agreed with that expected from SM production.
Searches~\cite{D0WW01,D0WW02} for an excess of $p\bar{p}\rightarrow WW+X$,
where the $W$ bosons each decayed to $\ell \nu$ $(\ell=e$ or $\mu )$,
yielded events which matched the SM prediction. Further, the $p_T$
spectrum of the charged leptons agreed~\cite{D0WW02} with the prediction.
Studies~\cite{D0WW03,D0WW04} of the processes $p\bar{p}\rightarrow WW + X$
and $p\bar{p}\rightarrow WZ + X$, where one $W$ boson decayed to an electron
or positron and the corresponding antineutrino or neutrino and the other
vector boson decayed to a quark-antiquark pair manifested as jets, yielded no
excess of events and a $W$ boson transverse energy spectrum which matched the
expected background plus SM signal.  Limits on
anomalous $WW\gamma$ and $WWZ$ couplings were derived from each of these
analyses.  Several~\cite{D0WG01,D0WW01,D0WW03} of these analyses were presented
in detail in Ref.~\cite{D01APRD}. The results of all of these analyses were
combined~\cite{D0WWWgWZ}, using the method described in Ref.~\cite{D01APRD}, to
form our most restrictive limits on anomalous $WW\gamma$ and $WWZ$ couplings.

Limits on the $WW\gamma$ couplings have been set by the UA2 and CDF
collaborations from the properties of $W+\gamma$ events~\cite{wgua2,wgcdf}
and by the L3 Collaboration~\cite{L3wev} from the rate of single 
$W$ boson production at $\sqrt{s} = 172$ GeV. Both 
the $WWZ$ and $WW\gamma$ couplings have been studied by several experiments.
CDF has searched for anomalous $WW$ and $WZ$ production~\cite{CDFWZ,CDFWW}
and the four experiments at the LEP $e^+ e^-$ collider 
have studied the properties of $WW$ events
~\cite{L397a,L397b,Op97,Op98,Op99,Del97,Del98,Ale98}.

In this paper two new analyses resulting from a
study of $p\bar{p}$ collisions at a center-of-mass energy of $\sqrt{s}=1.8$
TeV are presented. The collisions were recorded at D\O \ 
during the 1994--1995 and 1996 collider runs of the Fermilab Tevatron.

The first analysis is a search for $WZ$ production which provides a
test of anomalous couplings unique among the gauge boson pair analyses.
$WZ$ production is sensitive only to the $WWZ$ couplings, not the $WW\gamma$
couplings. In this analysis the collisions were searched for $WZ$ events where
the $Z$ boson decayed to $ee$ and the $W$ boson decayed to either 
$e\nu$ or $\mu\nu$.  The expected SM $WZ$ signal and the
background were approximately equal in size and both were expected to be small. 
The number of events observed was compared with that expected from anomalous
$WZ$ production in the presence of background to set upper limits on anomalous
$WWZ$ couplings.  

The second analysis is a search for anomalous $WW$ and $WZ$ 
$(WW/WZ)$ production, similar
to those of Refs.~\cite{D0WW03,D0WW04}, using the decay signature
$W\rightarrow\mu\nu$, $W/Z \rightarrow$ hadronic jets.
Because SM
$WW$ and $WZ$ production was swamped by backgrounds from other sources of
$\mu\nu {\rm jj}$ events, the analysis was sensitive only to anomalous
vector boson pair production. The $p_T$ spectrum of the $\mu\nu$
system was compared to that expected from anomalous $WW$ and $WZ$ production
plus the background, and limits on anomalous $WWZ$ and $WW\gamma$
couplings were produced.

The paper is arranged so that the subsequent
two sections present elements common
to the two analyses: the detector and particle identification. The fourth
section is a description of the $WZ\rightarrow \ell\ell\ell\nu +X$ search and
limits on anomalous $WWZ$ couplings.  The next section describes the
$WW/WZ \rightarrow \mu\nu {\rm jj}+X$ analysis and limits on anomalous
$WWZ$ and $WW\gamma$ couplings. The sixth section contains a summary of the
results of combining the anomalous coupling limits of these two analyses
with those of our previous publications, producing the most restrictive
anomalous $WW\gamma$ and $WWZ$ coupling limits available to date
from this experiment.  Finally, the last section contains 
the conclusion and summary of the results presented in this paper.

\section{Detector}
\begin{figure}
   \epsfxsize = 7.5cm
   \centerline{\epsffile{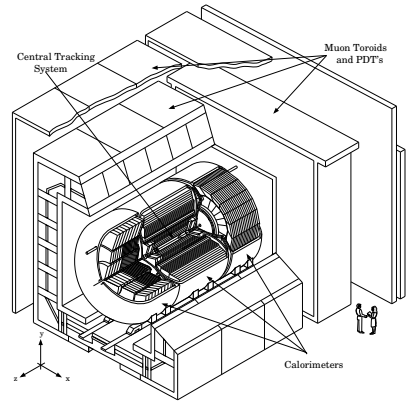}}
\caption{Isometric view of the D\O\ detector. Also shown are the calorimeter
support platform, the Tevatron beampipe centered within the calorimeter, and 
the Main Ring beampipe which penetrated
the muon system and calorimeter above the detector center.}
\label{fig-d0_det}
\end{figure}

The D\O \ detector consisted of four main systems: a non-magnetic 
inner tracking system, a liquid-argon uranium calorimeter, a muon 
spectrometer, and a trigger system.  The detector is briefly described 
in this section.  A detailed description of the detector is 
available in Ref.~\cite{d0nim}.  The tracker, calorimeter, and muon system 
are shown  in Fig.~\ref{fig-d0_det}.

A non-magnetic central tracking system, composed of central and
forward drift chambers, provided directional information for charged
particles and is used in this analysis to discriminate between
electrons and photons, and in muon identification.

Particle energies were measured by a liquid-argon uranium
sampling calorimeter that was divided into three cryostats.
The central calorimeter (CC) covered pseudorapidity~\cite{pseudo}
 $|\eta| < 1.1$, and the
end calorimeters (EC) covered $1.1 < |\eta| < 4.4$.  
The calorimeter was transversely segmented into projective towers with
$\Delta \eta \times \Delta \phi = 0.1\times 0.1$, where $\phi$ is the azimuthal
angle. The third layer of the electromagnetic (EM) calorimeters, where the
maximum energy deposition from EM showers was expected to occur, was
segmented more finely into cells with $\Delta \eta \times \Delta \phi =
0.05\times 0.05$. The scintillator-based 
intercryostat detectors (ICD's), which improved the energy resolution for 
jets that straddled the central and end calorimeters, were inserted into 
the space between the cryostats.  Thus, jet identification was performed 
in the whole calorimeter without any gap in pseudorapidity. 
Electron identification was performed for EM clusters with 
pseudorapidity $|\eta|\le 2.5$; but the boundary between the CC and EC 
cryostats resulted in a gap spanning the region $1.1\le |\eta| \le 1.5$. 
 
\begin{figure}
   \epsfxsize = 7.5cm
   \centerline{\epsffile{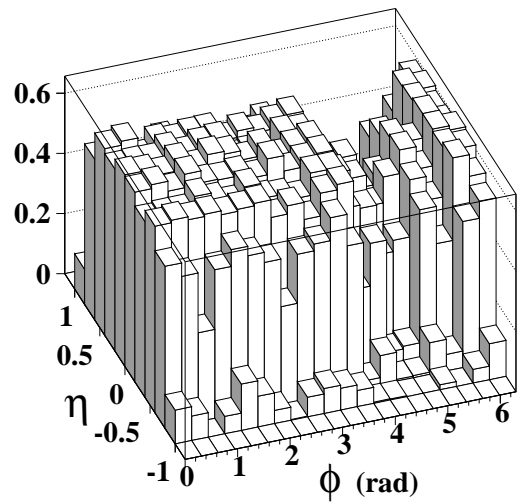}}
\caption{The geometrical acceptance of the muon detector within the 
region $|\eta|\le 1$. $\phi=\frac{3\pi}{2}$ is the downward $(-\hat{y})$
direction where the calorimeter support
 platform breaks into the muon system three-layer geometry. }
\label{fig-muaccept}
\end{figure}

The muon spectrometer  consisted of solid-iron toroidal magnets and
sets of proportional drift tubes (PDT's). It provided identification of muons
and determination of their trajectories and momenta. 
It consisted of three layers: a layer with four planes 
of PDT's, located between the calorimeter and the toroid magnets; and
two layers, each with three planes of PDT's,
located outside the toroid magnets.  Figure~\ref{fig-muaccept} shows the 
geometric acceptance of the muon detector for the region $|\eta|\le 1$ as
determined from a Monte Carlo simulation of the detector. 
The muon momentum $p$ was determined from its deflection angle in the magnetic
field of the toroid.  The momentum resolution was limited by multiple
scattering in the calorimeter and toroid, knowledge of the magnetic field
integral, and the accuracy of the deflection angle measurement.  

A multi-level, multi-detector trigger system~\cite{D01APRD,d0nim}
was used for selecting interesting events and recording them to tape.
A coincidence between hits in two hodoscopes
of scintillation counters (Level 0), centered around the beampipe,
was required  to register the presence of an inelastic collision.
These counters also served as the luminosity monitor for the experiment.
The Level 1 and Level 1.5 triggers were programmable hardware triggers 
which made decisions based on combinations of detector-specific algorithms.
The Level 2 trigger was a farm of 48 VAX 4000/60 and 4000/90 computers which 
filtered the events based on reconstruction of the information available 
from the front-end electronics. 

\section{Particle Identification}
The analyses described in this paper rely on the detector's ability 
to identify electrons, muons, hadronic jets, and the undetected transverse
energy due to neutrinos.  A brief description of the particle identification
criteria is presented in this section. A more detailed description of these 
particle identification criteria is available in Ref.~\cite{D01APRD}. 
 
\subsection{Electron Identification}
\label{sec-elecid}
Electron candidates were identified using information from the calorimeters and 
tracking detectors. Electron candidates were formed from clusters, identified 
using a nearest-neighbor algorithm, with more than $90\%$ of their energy 
in the EM layers of the calorimeter.  The EM
clusters had to fall within the CC ($|\eta|<1.0$) or 
either EC ($1.5<|\eta|<2.5$). Electrons  had to be 
isolated, had to have a shower shape consistent with that from test 
beam measurements, and had to have either a track that closely matched the
position of the shower centroid (``tight" selection criteria) or 
drift chamber  hits consistent with the passage 
of a charged  particle within an azimuthal road of width $\Delta\phi = 15 \;
(30)$ milliradians for CC (EC) EM clusters (``loose" selection criteria). 

The efficiency for selecting electrons with the selection criteria described 
above was calculated using $Z\to ee$ decays. The efficiencies for each 
$\eta$ region and electron definition are shown in Table~\ref{table.eideff}.
The energy resolution was  
$\sigma(E)/E=   14\%/\sqrt{E({\rm GeV})} \oplus 0.3\% \oplus 14\%/E({\rm GeV})$
for electrons in the CC and 
$\sigma(E)/E= 15.7\%/\sqrt{E({\rm GeV})} \oplus 0.3\% \oplus 29\%/E({\rm GeV})$ 
for electrons in the EC, where ``$\oplus$" indicates addition in 
quadrature.

\subsection{Muon Identification}
\label{sec-muid}
Muon candidates were tracks in the muon chambers which survived a
number of reconstruction quality cuts. A muon was required to lie within the 
central region ($|\eta|<1.0$). A muon  had to pass through a region
of the muon toroid with sufficient magnetic field 
($\int B dl>2.0$ Tesla-meters).  
The energy deposited along the muon track in the calorimeter had to be 
at least that expected from a minimum-ionizing particle
which on average deposits $\sim 1$ GeV.
The impact parameter of the muon with respect to the interaction
point had to be less than 20 cm. The muon track was refitted with the 
timing, $t_0$, of the muon track with respect to the collision as a floating 
parameter.  It was required that $t_0$ be consistent with a muon 
originating from the interaction. 
A slightly different $t_0$ cut was used in 
the two analyses due to the different nature of the backgrounds. 
Lastly, the muon  had to be separated by 
$\Delta R_{\mu}\equiv \sqrt{(\Delta\eta)^2+(\Delta\phi)^2}
\ge 0.5$ from the nearest jet in the event.  
The muon reconstruction efficiency in the $WZ\rightarrow {\mu \nu ee}$ 
($WW/WZ\rightarrow \mu\nu$jj) analysis for muons with $|\eta_{\mu}|<1$ was 
$0.701 \pm 0.031$ ($0.680^{+0.041}_{-0.080}$) excluding losses due to the 
geometric acceptance of the muon detector. The muon momentum resolution was 
$\sigma(\frac{1}{p})=0.18(p-2)/p^2\oplus 0.003$ ($p$ in GeV/$c$).

\subsection{Jet Identification and Missing Energy}
Jets were identified~\cite{D01APRD}
 as clusters of calorimeter towers within a cone centered 
on the highest $E_T$ tower. For the analyses described here, a cone size of 
$R\equiv \sqrt{(\Delta \eta)^2 + (\Delta \phi)^2} = 0.5$ was used. 
The energy deposited by the jet in the electromagnetic and
hadronic calorimeters  had to be consistent with that of an ordinary jet,
thus suppressing the backgrounds from isolated noisy calorimeter cells and
accelerator losses.  These jet identification criteria have an efficiency
of $0.96\pm0.01$ per jet. The jet energy resolution depended 
on the jet pseudorapidity and was approximately 
$\sigma(E)/E$ = $80\%/\sqrt{E({\rm GeV})}$. 

The primary sources of missing transverse energy 
included neutrinos, which escaped undetected, and the energy 
imbalance due to the
resolution of the calorimeter and muon system.
Two calculations of missing transverse energy 
were made. The missing transverse energy which was calculated from the 
energy deposited in the calorimeter is referred to as 
\hbox{$\rlap{\kern0.25em/}E_T^{{\rm cal}}$}. The missing energy which was 
calculated from the energy deposited in the calorimeter and was corrected for 
muons passing some loose quality cuts is referred to as 
\hbox{$\rlap{\kern0.25em/}E_T$}.

\section{Search for $WZ\rightarrow$ Trileptons}
A search for $WZ$ production was performed in the $e\nu ee$ and $\mu \nu ee$
decay modes, taking advantage of the unusual signature consisting of three 
charged high-$E_T$ leptons and the missing transverse energy due to the 
high-$E_T$ neutrino.

\label{sec-trilep}
\subsection{Trigger and Data Sample}
The Level 1 trigger used for this study required two EM calorimeter trigger 
towers $(\Delta \eta \times \Delta \phi = 0.2 \times 0.2)$
with $E_T>10$ GeV. The Level 2 trigger required two clusters of EM trigger 
towers which had  $E_T>20$ GeV and passed Level 2 isolation and shower shape 
cuts. The efficiency of the trigger was measured as a function of the 
reconstructed electron $E_T$ and found to be greater than 99\% for a 
reconstructed $E_T>25$ GeV. The integrated luminosity of the data sample was 
$92.3\pm5.0$ pb$^{-1}$. The luminosity determination is described in 
Ref.~\cite{d0lum}.

\subsection{Event Selection Criteria}
$WZ\rightarrow e\nu ee$ events were required to have two high-$E_T$
electrons consistent with a $Z$ boson decay, and a third electron and
\hbox{$\rlap{\kern0.25em/}E_T$} consistent with a $W$ boson decay.
Specifically, at least one electron was required to satisfy the tight
selection criteria and another two were required to satisfy the tight
or loose selection criteria (as defined in Section~\ref{sec-elecid}). A tight 
electron and one of the other electrons were required to have 
$E_T>25$ GeV and the third electron to have $E_T>10$ GeV.
It was required that \hbox{$\rlap{\kern0.25em/}E_T^{{\rm cal}}$} $>15$ GeV. The 
invariant mass of two of the electrons had to be within the range  
$81<M_{i,j}<101$ GeV/$c^2$, as expected for the decay of a $Z$ boson.  The 
transverse mass $$ M_T(e\nu) = \sqrt{2E^e_T
 \hbox{$\rlap{\kern0.25em/}E_T^{{\rm cal}}$}(1-\cos(\phi_e-\phi_\nu))}$$
calculated using the $E_T$ of the other electron and the 
\hbox{$\rlap{\kern0.25em/}E_T^{{\rm cal}}$} 
was required to be $M_T(e \nu)>30$ GeV,
as expected for the decay of a $W$ boson.  These criteria were checked for all
three combinations of electrons.  One event was found which passed all 
the selection criteria. The parameters~\cite{patsthesis}
of this event are described in the appendix.

$WZ\rightarrow \mu\nu ee$ events were required to have two high-$E_T$
electrons as expected for a $Z$ boson decay, and a muon and
\hbox{$\rlap{\kern0.25em/}E_T$} consistent with a $W$ boson decay.
Specifically, at least one electron was required to satisfy the tight
selection criteria and another was required to satisfy the tight
or loose selection criteria.
Both electrons had to have  $E_T>25$ GeV. 
Instead of the $10$ GeV third electron of the $e\nu ee$ search, 
a muon with $p_T>15$ GeV/$c$ was required.  Finally, it was required that 
\hbox{$\rlap{\kern0.25em/}E_T$} $>15$ GeV. No events passed these selection 
criteria.

\subsection{Background Expected}
The trilepton plus missing transverse energy signature demanded by the
event selection has no known significant sources other than $WZ$ production 
and backgrounds due to objects misidentified as leptons. 
 
In the $e\nu ee$ channel the largest background was expected to come from 
$Z$ + jet events with $Z\to ee$ and where a jet mimicked an additional 
electron.  This background was estimated using  data.  
 Events with two electron candidates and one or more jets were 
selected from the  same data sample used in the event selection. The kinematic
event selection  criteria were applied treating each jet as the third electron. 
The probability for a jet to mimic a tight or loose electron was determined
from a sample of multijet events and was 
parameterized by a linear function of jet
$E_T$ for jets with $E_T$ less than $\sim 150$ GeV, as given in 
Table~\ref{table-pje}. The background was then the number of 
$ee + {\rm jet}$ events times the probability of a jet 
mimicking the third (tight or loose) electron.  This background
was estimated to be $0.38~\pm~0.07({\rm stat})~\pm~0.11$({\rm syst}) events. 
The size of the  statistical uncertainty was determined by the statistics 
of  the $ee + $ jets sample. 
The systematic uncertainty was dominated by the 25\%
uncertainty in the probability for a jet to mimic a tight or loose
electron. This latter uncertainty was due, in large part, to the 
uncertainty on the amount of direct photons in the multijet
sample. A cross check based on
a data sample of events enriched with highly EM jets which failed
the electron selection criteria gave $0.42^{+0.41}_{-0.26}$ events for
this background.

In the $\mu\nu ee$ channel there were two contributions to the background, 
one from events with two electrons and a jet which produced an isolated muon 
and one from events with an electron, a muon, and a jet which mimicked an 
electron. Data-based methods of calculating the background were used to 
estimate both of these contributions to the background. 

To calculate the $ee + {\rm jet}$ event background, events with two
electrons and a central jet were selected (this was called the 
``fake" sample). 
Each event was required to pass all selection criteria except that the jet
was only required to pass the muon fiducial and kinematic selection. The
number of events was then multiplied by the probability of the jet
producing an isolated muon, this probability having been determined using
two methods.  The probability (per jet) of finding an
isolated muon in a sample of multijet events with $E_T({\rm jet}) > 15$ GeV
was found to be $1.5\times10^{-5}$.  
The number of  events expected from this background was $\le 0.002$. 
On the other hand, a fraction of the $ee+{\rm jet}$ 
events contained heavy quark 
($b/c$) jets. Assuming that all of the jets in the fake sample are heavy 
quark jets, 
a heavy-quark-enhanced fake rate was used to obtain an upper limit for this 
background. The probability of a jet mimicking a muon from a heavy quark 
($b/c$) jet was found by requiring a muon (isolated or non-isolated) in the 
opposite  hemisphere from the isolated muon in multijet events. This gave a 
heavy-quark-enhanced fake rate of $2.5\times 10^{-4}$,
resulting in an upper limit of $N_{{\rm bkg}}=0.022\pm0.004$ events. When
setting limits on the cross section and coupling parameters, a smaller
background estimate gives a more conservative limit. Therefore the 
lower estimate ($\le 0.002$ events) was used in lieu of the larger  
($0.022$ events).  

The second background ($e\mu + {\rm jet}$ events) was calculated using events 
collected with a different trigger 
which required one EM object with $E_T>20$ GeV and 
\hbox{$\rlap{\kern0.25em/}E_T$}$> 20$ GeV.
Events were selected which had an isolated muon, one or more jets, and a 
tight or loose electron. All event selection cuts were applied with the 
exception of the trigger. The number of background events was then found by
summing the $E_T$-dependent probability for a jet to have mimicked an 
electron for each event which passed the event selection criteria, 
accounting correctly for  events which contained more than one 
jet and the difference in the integrated luminosities between the two triggers. 
The total number of background
events expected from $e\mu +$ jet events 
was found to be 0.118~$\pm$~ 0.018(stat)~$\pm$~0.035(syst).  
Again, the systematic error is due to the uncertainty in the
probability for a jet to mimic an electron.

The total background to $WZ\rightarrow \rm{ trileptons} $ was $0.50\pm
0.17$ events.

\subsection{Efficiency Estimate}
The efficiency estimate was made using a leading-order Monte Carlo (MC) event 
generator\cite{HWZ} which also could simulate the effects of anomalous 
couplings. The \break 
MRSD-$^{\prime}$ parton distribution functions~\cite{mrs} were used. To 
correct for the effects of higher-order QCD processes which contribute to $WZ$
production, the resulting cross section was increased by a $k$-factor of 
1.34~\cite{HWZ} and the $WZ$ system was given a transverse boost according 
to the  distribution produced by the {\sc pythia} Monte Carlo~\cite{pythia}
simulation of SM 
$WZ$ production. A parameterized detector simulation was used to account for 
the acceptance of the detector, the effects of detector resolution on the 
measurements of charged leptons and \hbox{$\rlap{\kern0.25em/}E_T$}, and 
the length $(\sigma \sim 30$ cm) of the $p\bar{p}$ collision region along 
the beam direction. 

For SM $WZ$ production, the detection efficiency in the $e\nu ee$ and  
$\mu\nu ee$ channels was found to be $(16.9\pm 1.4)\%$ and $(11.5\pm 1.4)\%$,
respectively.  For a SM cross section of
$2.6$ pb~\cite{ohnwz}, the $( W\rightarrow \ell \nu ) \times
 (Z\rightarrow ee)$ branching fractions~\cite{pdg}, and an integrated
luminosity of $92.3\pm 5.0$ pb$^{-1}$, the expected number of events in the
$eee\nu$ and $ee\mu\nu$ channels was 
$0.146\pm 0.002({\rm stat})\pm 0.012({\rm syst})$ and
$0.099\pm 0.001{\rm (stat)} \pm 0.009{\rm (syst)}$, respectively.
The small statistical uncertainties reflected the large number of 
MC events generated and processed through the detector simulation.
The systematic error included the uncertainties in the luminosity (5.3\%), 
the particle  identification efficiency (0.7\%), the trigger efficiency (2\%), 
the branching fraction (3.7\%), and the 
MC cross section due to the choice of the parton 
distribution function and $Q^2$ scale (5\%). 
The total expected signal from SM $WZ$ production was $0.25\pm 0.02$ events.
The results are summarized in Table~\ref{results}.

\subsection{$WZ$ Production Cross Section Limit}
The $95\%$ confidence level (C.L.) 
upper limit on the $WZ$ cross section is estimated
based on one observed event and a subtraction of the expected background
of $0.50\pm 0.17$ events.  Poisson-distributed numbers of events were 
convoluted with Gaussian uncertainties in the efficiency and background. 
For $WZ$ production, the $95\%$
C.L. upper limit on the cross section was 47~pb, consistent with, but much 
larger than, the SM prediction.

\subsection{Limits on Anomalous $WWZ$ Couplings}
The event generator~\cite{HWZ} and parameterized detector simulation were
used, in a manner identical to that described above, to find the efficiency 
and expected number of events in the case of hypothetical anomalous $WZ$ 
couplings. A grid in the $\lambda_Z$--$\Delta g_1^Z$ plane was used.   
Once the probability for observing one event was determined~\cite{patsthesis}
for each point in the grid, limits on the anomalous couplings were made. 
The limits were found by taking the logarithm of the likelihood 
and identifying the contour in $\lambda_Z - \Delta g_1^Z$ 
around the point of maximum of the logarithm of the likelihood 
$(L_{{\rm max}})$  where $L=L_{{\rm max}}-\delta$. 
To set a 95\% C.L. limit in one dimension, the contour was 
evaluated at $\delta=1.92$. To set a 95\% C.L. limit in
two dimensions (allowing two anomalous couplings to vary at the same time), 
the contour was evaluated at $\delta=3.00$.

\begin{figure}
   \epsfxsize = 7.5cm
   \centerline{\epsffile{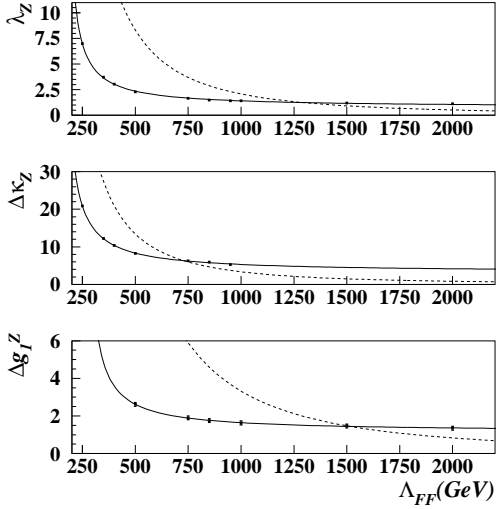}}
\caption{One-dimensional 95\% C.L. (solid) and unitarity limits (dashed) vs.
$\Lambda$ for the $WWZ$ coupling parameters $\lambda _{Z}$,
$\Delta\kappa_Z$, and $\Delta g^Z_1$.}
\label{bounds_eee}
\end{figure}

The value of the form factor scale $\Lambda$
was chosen such that the coupling limit was less than the unitarity 
limit~\cite{baurpc}. 
The one-dimensional 95\% C.L. coupling limits and unitarity
limits as a function of $\Lambda$ for each of the three 
coupling parameters are shown in Fig.~\ref{bounds_eee}.

\begin{figure}
   \epsfxsize = 7.5cm
   \centerline{\epsffile{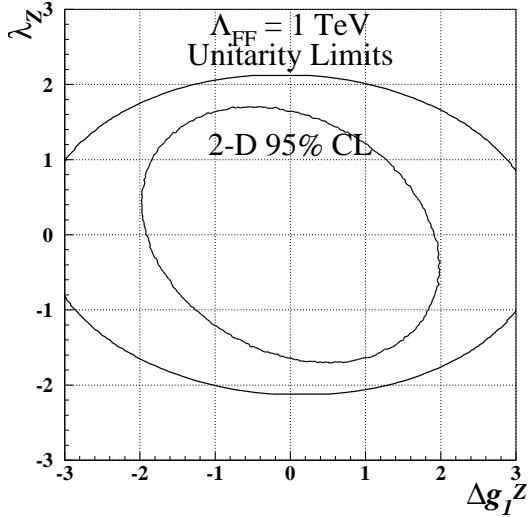}}
\caption{Correlated limits on $\Delta g^{Z}_1$
and $\lambda _{Z}$ for $\Lambda$ = 1 TeV obtained from a fit to the cross
section using the 1994--1996 data 
for the $\mu\nu ee$ and $e\nu ee$ channels combined.
The inner solid line is the two-dimensional 
95\% C.L. limit and the outer solid  line is the
unitarity limit.}
\label{zw_lim_1000}
\end{figure}

This analysis was most sensitive to the parameters $\lambda_Z$ and $\Delta
g_1^Z$. Setting $\Lambda=1$ TeV, the one-dimensional 95\% C.L. limits from 
the $e\nu ee$ and $\mu\nu ee$ channels are  
\[|\Delta g_1^Z|<1.63\]
\[|\lambda_Z|<1.42 \]
\noindent when all other parameters are held at their SM values.
The two-dimensional 95\% C.L. contour limits for $\Lambda=1$ TeV are shown in
Fig.~\ref{zw_lim_1000} for the $e\nu ee$ and $\mu\nu ee$ data combined. 

\section{Search for Anomalous $WW$ and $WZ$ Production}
\label{sec-mujets}
The 1994--1995   data  were searched 
for anomalous $WW/WZ$ production in events with the signature: high-$p_T$ muon;
large \hbox{$\rlap{\kern0.25em/}E_T$}; and at least two jets 
($\mu\nu {\rm jj})$. 

\subsection{Trigger and Data Sample}
The  Level 1 trigger consisted of a muon candidate in the central 
region and at least 5 GeV deposited in a hadronic trigger 
tower $(\Delta \eta \times \Delta \phi = 0.2 \times 0.2)$.
As the muon scintillation counters became available during the 
collider run they were added to the Level 1 trigger in such a way as to veto 
out-of-time muons, such as those that originated from cosmic rays.  

\begin{figure}
   \vspace*{-1.in}
   \epsfxsize = 7.5cm
   \centerline{\epsffile{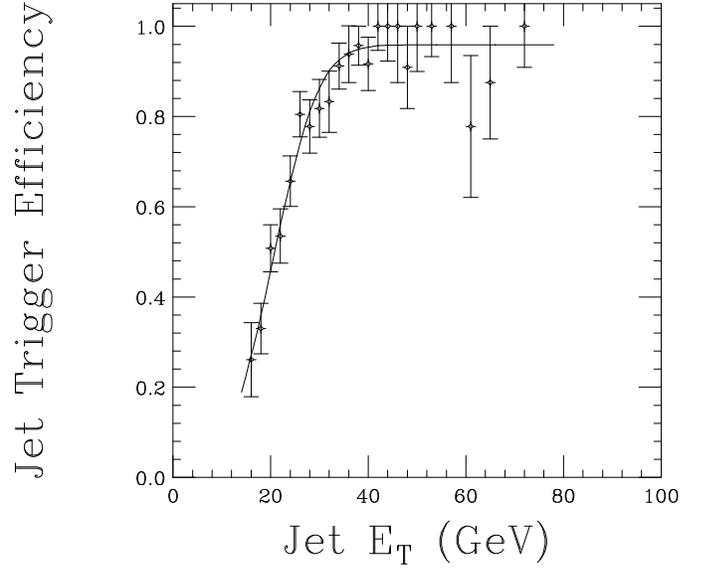}}
   \vspace*{1.in}
\caption{Jet trigger efficiency in pseudorapidity region $|\eta|<1.0$.
The curve is the result of an error function fit to the efficiency.}
\label{fig-trig_10}
\end{figure}

The Level 2  trigger required a muon with $p_T>10$ GeV/$c$, as determined by 
the muon pattern recognition algorithm taken from the reconstruction program.
A jet was required with $E_T>15$ GeV within the region $|\eta|<2.5$. 
The jets were identified by a cone algorithm which summed $E_T$'s of 
calorimeter towers in cones of $R = 0.7$.  
The efficiency of the jet part of the Level 1 and Level 2 
triggers was measured as a function of the reconstructed jet $E_T$ in three 
separate pseudorapidity bins by comparing the results of the single-muon trigger
with the single-muon plus jet trigger for events which contained a single jet.
Figure~\ref{fig-trig_10} shows the jet trigger efficiency as a function 
of jet $E_T$ for the pseudorapidity region $|\eta|<1.0$. The jet trigger 
efficiency reached a plateau at jet $E_T$ of approximately 40 GeV. 
The efficiency was parameterized using an error function.  The curve shown
in Fig.~\ref{fig-trig_10} is the result of that fit.  
The results in the other two 
pseudorapidity regions were similar.   For SM Monte Carlo events which passed
all of the selection criteria, the efficiency of the jet part of the trigger 
was $0.927\pm0.007$.  An alternate fit with a plateau at
$100\%$ increased this efficiency by $0.012$ and that was taken as a systematic
uncertainty.  The efficiency of the muon component of the trigger was 
$0.707\pm0.018$. 
The integrated luminosity~\cite{d0lum} 
of the data sample was $80.7\pm 4.3$ pb$^{-1}$.

\subsection{Event Selection Criteria}
\begin{figure}[htb]
   \hspace*{-0.04in}
   \epsfxsize = 7.5cm
   \centerline{\epsffile{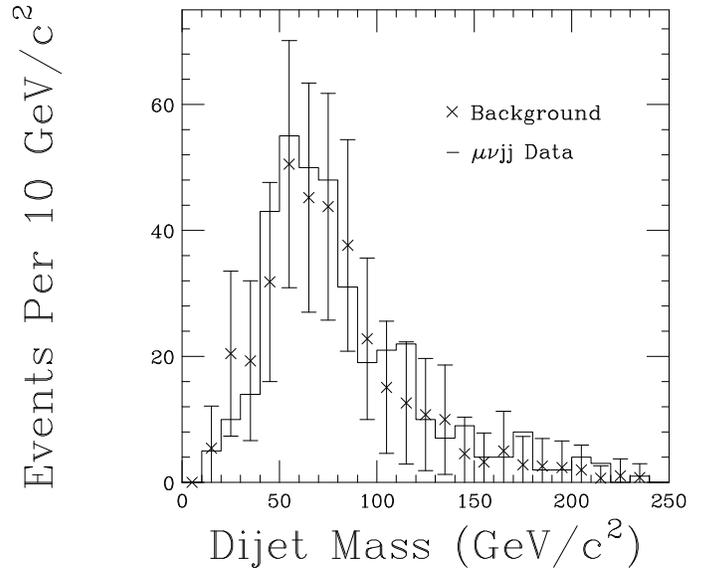}}
   \vspace*{0.5in}
\caption{Comparison of invariant mass of the two highest $E_T$ jets for 
the data (histogram) and the estimated total background (points with 
uncertainties) for the $WW/WZ\rightarrow \mu\nu {\rm jj}$ channel. 
The uncertainties shown are statistical only.}
\label{fig-dijetmass}
\end{figure}

The signature of the muon + jets channel consisted of an isolated high-$p_T$ 
muon from the $W$ boson decay and a minimum of two jets from a
$W$ or $Z$ boson decay. We did not differentiate between the two processes 
$W \to {\rm jj}$ and $Z \to {\rm jj}$ due to the dijet mass resolution of
the calorimeter.  Single muon events with the following 
characteristics were selected.  The muon was within the central region, 
which corresponded
approximately to $|\eta|<1$, and had transverse momentum 
$p_{T}^{\mu} \ge 20$ GeV/$c$.  A  \hbox{$\rlap{\kern0.25em/}E_T$} 
of at least 20 GeV was required in each event.  Demanding a transverse 
mass $M_T(\mu\nu)>$ 40 GeV/$c^2$, where
$$M_T(\mu\nu) = \sqrt{2E^\mu_T
\hbox{$\rlap{\kern0.25em/}E_T$}(1-\cos(\phi_\mu-\phi_\nu))},$$ 
completed the kinematic selection defining the decay of a $W$ boson
candidate.  Next, the candidates had to contain
at least two jets ($|\eta|<2.5$) with $E_T \ge 20$ GeV.  
The invariant mass of the two highest $E_T$ jets  had to be between 
50 and 110 GeV/$c^2$ as expected for the decay of a $W$ or $Z$ boson.
Figure~\ref{fig-dijetmass} displays the distribution of the 
invariant mass of the two highest 
$E_T$ jets in the 372 events which remained in the sample after all selection 
criteria, except for the dijet mass selection, had been applied. 

\begin{figure}[htb]
   \vspace*{-1.in}
   \hspace*{-0.04in}
   \epsfxsize = 7.5cm
    \centerline{\epsffile{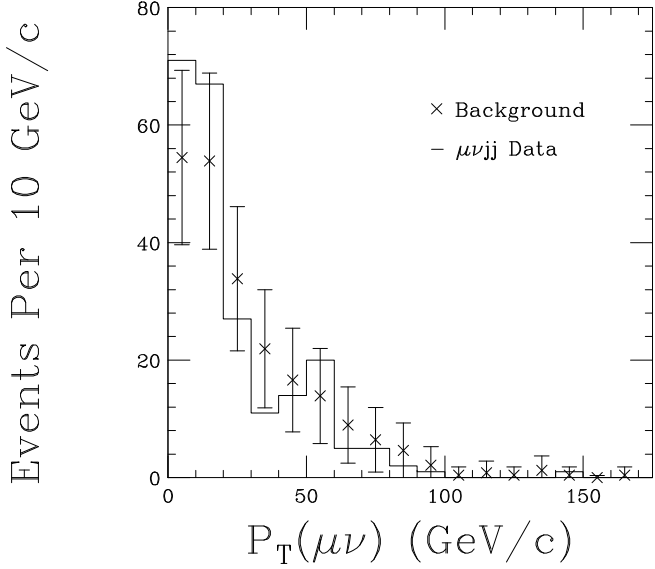}}
\vspace*{0.75in}
\caption{Comparison of the $p_T(W)$ distributions of signal (histogram) 
and estimated total background ($\times$ with statistical uncertainties) 
for $WW/WZ \rightarrow \mu\nu {\rm jj}$. They are
consistent with each other indicating the presence of no significant
anomalous gauge couplings. }
\label{fig-dabkg}
\end{figure}

Application of the above cuts led to a final data sample of 224
events. The $p_T(\mu\nu)$ distribution for these events is shown in 
Fig.~\ref{fig-dabkg}. The distribution indicates absence of events at
$p_T(\mu\nu){\ }>{\ }150$ GeV/$c$.  The $W$ boson candidate 
with the highest transverse  energy had $p_T(\mu\nu)=141$ GeV/$c$. 

\subsection{Background Expected}
There were two major sources of background to the $WW/WZ \to \mu\nu
{\rm jj}$ production: $W{\ }+\ge{\ }2$ jets with $W \to \mu\nu$; and
 QCD multijet events where one of the jets was accompanied by a muon 
which was misidentified as an isolated muon and where there was significant 
\hbox{$\rlap{\kern0.25em/}E_T$}. The latter background could have arisen 
from $b$-quark pair production, for instance. 
Contributions from other backgrounds such as: $t{\overline
t}$ production with subsequent decay to $W^+bW^-{\overline b}$
followed by $W \to \mu\nu$; $WW/WZ$ production with $W \to \tau\nu$ 
followed by $\tau \to \mu\nu{\overline \nu}$; 
$ZX \to \mu\mu X$, where one of the muons 
was missing; and $ZX \to \tau\tau X$ with $\tau \to \mu\nu{\overline
\nu}$, were small or negligible. 

The QCD multijet background was estimated using a background 
enriched data sample. This technique was similar to that used in our 
previous analysis~\cite{topprd}. 
The probability for a jet with a muon to be misidentified
as an isolated muon was determined from the ratio of the number of events 
containing an isolated muon, at least one jet, and 
\hbox{$\rlap{\kern0.25em/}E_T$} less than 20 GeV to the number of events 
which contained a muon which failed the jet isolation requirement (but 
otherwise passed the muon identification cuts), two or more jets, and 
\hbox{$\rlap{\kern0.25em/}E_T$} less than 20 GeV.  This probability was 
$0.041\pm0.007$.  Then the number of events which passed all of the 
selection criteria of the signal except for the muon-jet isolation 
requirement, again applied in reverse so as to form a sample 
complementary to the signal, was counted.  
This provided the sample
of events for which misidentification of a non-isolated muon as an isolated 
muon would have created a false signal.  There were 2567 such events.  Thus the
QCD multijet background was $105\pm19$ (stat) events. 
  The QCD multijet background was also calculated for
events which passed all the selection criteria for the signal except for
the dijet invariant mass selection, which was applied in reverse.  This 
number was necessary for performing a background subtraction to the data 
in the out-of-mass cut region in order to calculate a normalization factor 
for the $W+ \ge {\rm 2}$ jets background. The QCD multijet background in the 
out-of-mass cut region was $55\pm14$ (stat) events. 

   The $W{\ }+\ge{\ }2$ jets background was estimated using the 
{\sc vecbos}~\cite{faber} event generator, with $Q^2 = (p^j_T)^2$, followed by
parton fragmentation using the {\sc herwig}~\cite{gmarc} package and a
detailed {\sc geant}-based~\cite{fcarm}  simulation of the
detector.  Normalization of the $W +\ge{\ } 2$ jets background was
determined by comparing the number of events expected from the {\sc vecbos}
estimate to the number of candidate events outside the dijet mass
window, after the QCD multijet contribution had been subtracted. 
The contribution from this background was calculated to
be $117\pm 24$ (stat) events.  A small component of the 
background, due to $Z{\ }+\ge{\ }2$ jets with an unreconstructed 
muon which mimicked the \hbox{$\rlap{\kern0.25em/}E_T$}, was accounted for in 
this procedure because of the kinematic similarity to $W$ boson decay. 
 
Among the other backgrounds, the only non-negligible contribution arose
from $t{\overline t} \to W^+bW^-{\overline b}$ decays. This was
estimated using a Monte Carlo sample  produced similarly to that of the $ W{\ }
+\ge$ 2 jets background sample. 
The $t\bar{t}$ background, calculated assuming a cross section of $5.5\pm1.8$
pb~\cite{d0topxs}, amounted to $2.7\pm1.2$  events.

The total expected background was $224\pm 31$ (stat) events.  The number
of observed events (224) was consistent with the background, and was much
larger than the predicted SM WW/WZ signal (discussed in the next section).
The systematic uncertainties in the QCD multijet background and the
$W + \ge 2$ jets background were correlated because of the common
uncertainty in the jet energy scale and because of the background
subtraction carried out in the normalization procedure when the
$W + \ge 2$ jets background was determined.  As a cross check,
the consistency between the background estimate and the number of observed
events was verified for variations of the event selection criteria.
The systematic uncertainties for the background estimation were:
dijet mass window selection (13.4\%); muon isolation (11.7\%); jet
energy scale (7.8\%); missing transverse energy selection (7.2\%);
and $W$ boson transverse mass selection (4.3\%).  The total
systematic uncertainty in the background was 46 events and
the total uncertainty in the background was 56 events.

The contributions from all background sources are shown in
Table~\ref{tab-evsum}.  The estimates in the table for the components of the
background include statistical uncertainties only.
Figure~\ref{fig-dijetmass} also displays the invariant mass of the two
highest-$E_T$ jets from the expected background with all selection
criteria, except for the dijet invariant mass selection, applied.
The final distributions of the signal and the sum of backgrounds are
plotted as a function of $p_T(W)$ in Fig.~\ref{fig-dabkg}.

\subsection{$WW/WZ$ Signal Estimate}
The efficiency for detecting $WW$ and $WZ$ events, for both SM and 
anomalous couplings, was determined using a
leading-order event generator~\cite{HWZ} and a parameterized simulation 
of the detector.  The MRSD-$^{\prime}$ parton distributions~\cite{mrs} and 
a $k$-factor of 1.34~\cite{HWZ} were used in estimating the $WW/WZ$ 
cross section.  In order to simulate the kinematics associated with 
higher-order production processes, the diboson decay products were boosted in
the direction opposite to the hadronic recoil according to the $E_T$ 
distribution provided by {\sc pythia}~\cite{pythia} for SM $WW$ production.  
The efficiency was $2.5\%$ lower when this boost was turned off, and half 
of this difference was taken as the fractional systematic uncertainty.  
The interaction points were selected around the center of the nominal
collision point (z=0) from a Gaussian distribution with $\sigma =30$ cm. 

The muon fiducial acceptance was determined from a 
{\sc geant}-based~\cite{fcarm} detector model and is shown in 
Fig.~\ref{fig-muaccept}. 

\begin{figure}[htb]
   \vspace*{-1.in}
   \epsfxsize = 7.5cm
   \centerline{\epsffile{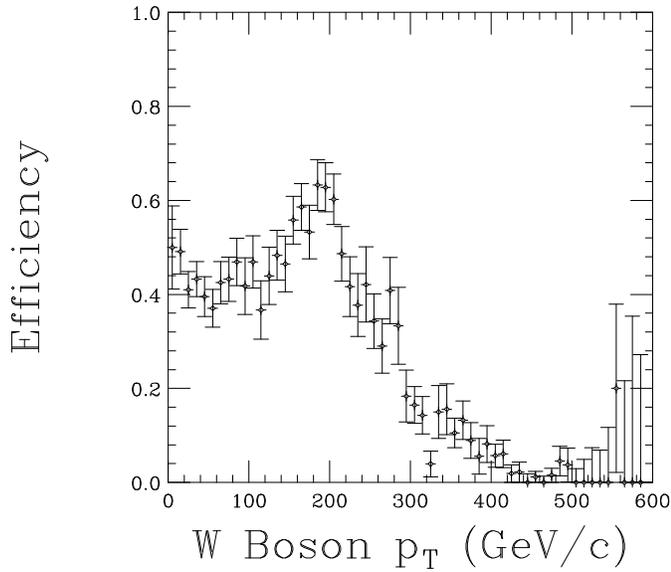}}
   \vspace*{1.in}
\caption{The efficiency of the dijet reconstruction and selection as a 
function of $p_T(\mu\nu)$ in the $WW/WZ\rightarrow \mu\nu$jj analysis.  
The uncertainties shown are statistical only. }
\label{fig-jetseffy2}
\end{figure}

The jets from a high-$p_T$ $W$ or $Z$ boson decay may have been close enough 
to  overlap and have poorly reconstructed energies, or they may have been 
completely merged into one jet.  Therefore, the efficiencies of the jet 
selection and  dijet mass selection depended on the boson's $p_T$.  
SM $WW$   events, generated using {\sc pythia} Monte Carlo and the
{\sc geant}-based detector model, were used to determine this efficiency 
as a function of $p_T(\mu\nu)$. The results were incorporated 
into the parameterized detector simulation.  Figure~\ref{fig-jetseffy2}
shows the efficiency as a function of $p_T(\mu\nu)$ for events which passed the 
rest of the event selection criteria.  The efficiency was low for low-$p_T$
$W$ boson events because of the jet $E_T$ threshold of $20$ GeV.  
It peaked at $63\%$ for  
$p_T(\mu\nu) = 200$ GeV/$c$ and fell for higher $p_T$ because of jet merging.
The uncertainty in the jet energy scale corrections led to a systematic 
uncertainty in the efficiency for $W$ and $Z$ boson identification of $3\%$. 

The kinematic efficiencies for SM $WW$ and $WZ$ detection were 
$0.073\pm0.002(\rm stat)\pm0.003(\rm syst)$ and 
$0.067\pm0.002(\rm stat)\pm0.010(\rm syst)$, respectively,  where the
additional systematic uncertainty originates from differences between 
the acceptances calculated with the parameterized detector simulation 
and the acceptances calculated using {\sc pythia} and {\sc geant} 
due to the jet reconstruction efficiency parameterization.  
Folding in the uncertainties due to the model of the jet trigger, the 
jet energy scale, and in the initial diboson boost, the systematic 
uncertainties in the kinematic efficiency amounted to $6.7\%$ and $15.8\%$ 
of the $WW$ and $WZ$ detection efficiency. 
Thus, the total efficiencies for SM $WW$ and $WZ$ production were 
$0.0351^{+0.0033}_{-0.0048}$ and $0.0322^{+0.0055}_{-0.0064}$, respectively.
The efficiency was slightly higher for simulated $WW$ and $WZ$ production  
with anomalous $WW\gamma$ and/or $WWZ$ couplings because the 
bosons originated at higher average $p_T$.  For instance, 
for $WW$ events produced with $\Lambda=2.0$ TeV, the total efficiency 
was $0.038^{+0.004}_{-0.005}$ for the case $\lambda=1.0$ and 
$\Delta\kappa=0.0$, and $0.043^{+0.004}_{-0.006}$ for the case 
$\lambda=2.0$ and $\Delta\kappa=2.0$.  

The predicted cross section~\cite{HWZ} for SM $WW$ $(WZ)$ production is 
$10.1$ $(2.6)$ pb.  A $5\%$ systematic uncertainty in this originates from the 
variation of the cross section depending on the set of parton distributions 
used in the event generation.  
The branching fractions~\cite{pdg} for $W\rightarrow \mu\nu$ and $W\rightarrow
{\rm jets}$ or  $Z\rightarrow {\rm jets}$ lead to overall branching fractions 
of $0.1412\pm0.0086$ and $0.0727\pm0.0042$, respectively.  
Therefore, with an integrated luminosity of $80.7\pm4.3$ pb$^{-1}$, 
$4.04^{+0.54}_{-0.68}$ $WW$ events and
$0.49^{+0.10}_{-0.11}$ $WZ$ events were expected to have been detected if 
production is solely through SM processes.  

\subsection{Limits on Anomalous $WW\gamma$ and $WWZ$ Couplings}
Since no excess of events in the high-$p_T(W)$ region was observed,
significant deviations from the SM trilinear gauge couplings were
excluded. Using the detection efficiencies for SM $WW$ and $WZ$
production and the background subtracted data, 
upper limits were set on the anomalous coupling parameters $\lambda$ and
$\Delta\kappa$. This determination was made using a binned likelihood fit 
of the observed $p_T(W)$ spectrum to the prediction of the Monte Carlo signal 
plus the estimated background. Unequal width bins were used to evenly
distribute the observed events, especially those in the high $p_T(W)$
region. In each $p_T$ bin for a given set of anomalous coupling
parameters, the probability for the sum of the background
estimate and Monte Carlo $WW/WZ$ prediction to fluctuate to the
observed number of events was calculated. The uncertainties in the background 
estimations, efficiencies, integrated luminosity, and Monte Carlo 
signal modelling were convoluted into the likelihood function using
Gaussian distributions. 

\begin{figure}[htb]
   \hspace*{0.03in}
   \epsfxsize = 7.5cm
 \centerline{\epsffile{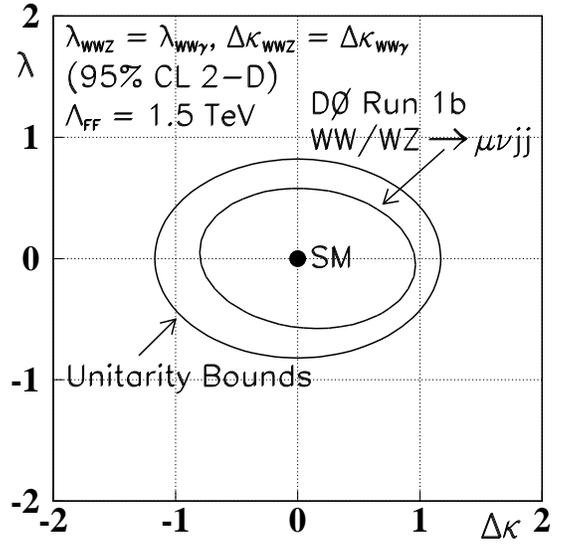}}
\caption{Contour plot of allowed region in the $\lambda-\Delta\kappa$
space for $WW/WZ \rightarrow \mu\nu {\rm jj}$ at 95\% C.L. for
$\Lambda=1.5$ TeV. The outer ellipse shows the bounds imposed
by the unitary relations on $\lambda$ and $\Delta\kappa$.}
\label{fig-limit}
\end{figure}

The one-dimensional 95\% C.L. limits on $\lambda$ and $\Delta\kappa$ 
are summarized in 
Table~\ref{tab-limits} for $\Lambda = 1.5$ TeV and 2.0 TeV. The first two
rows provide the coupling limits in the case of equal couplings for $WWZ$ and 
$WW\gamma$. The last two rows provide limits in the case of HISZ
relations~\cite{HISZ}. In each case, one of the couplings was fixed to
its SM value while the other was varied. The two-dimensional 
bounds (corresponding to a logarithm of the likelihood function
value 3.00 below the maximum value) for anomalous coupling parameters
in the $\lambda-\Delta\kappa$ plane are shown in Fig.~\ref{fig-limit}
for $\Lambda$ = 1.5 TeV.  Figure~\ref{fig-limit} also
shows the bounds imposed by the unitary conditions as a larger ellipse.

\section{Combined Results}
The results of the two searches described in this paper have been combined with 
those of our previous publications using the procedure described in 
Ref.~\cite{D01APRD}.  The method was to perform a binned maximum
likelihood fit of the number of events and their kinematic characteristics 
to the expected signals and backgrounds, taking care to account for correlated 
uncertainties among the data sets.  The number of events and the
expected background in the $WZ\rightarrow {\rm trileptons}$ analysis
of Section~IV, and the $p_T(\mu\nu)$ spectrum 
as well as the
expected background in the $WW/WZ$ analysis of Section~V, 
were included into the multiple final state fit described in 
Ref.~\cite{D0WWWgWZ}. The resulting limits on anomalous couplings 
represent the most restrictive available from our experiment. 

\begin{figure}[htb]
\hbox{
\epsfxsize = 1.7in
\epsffile{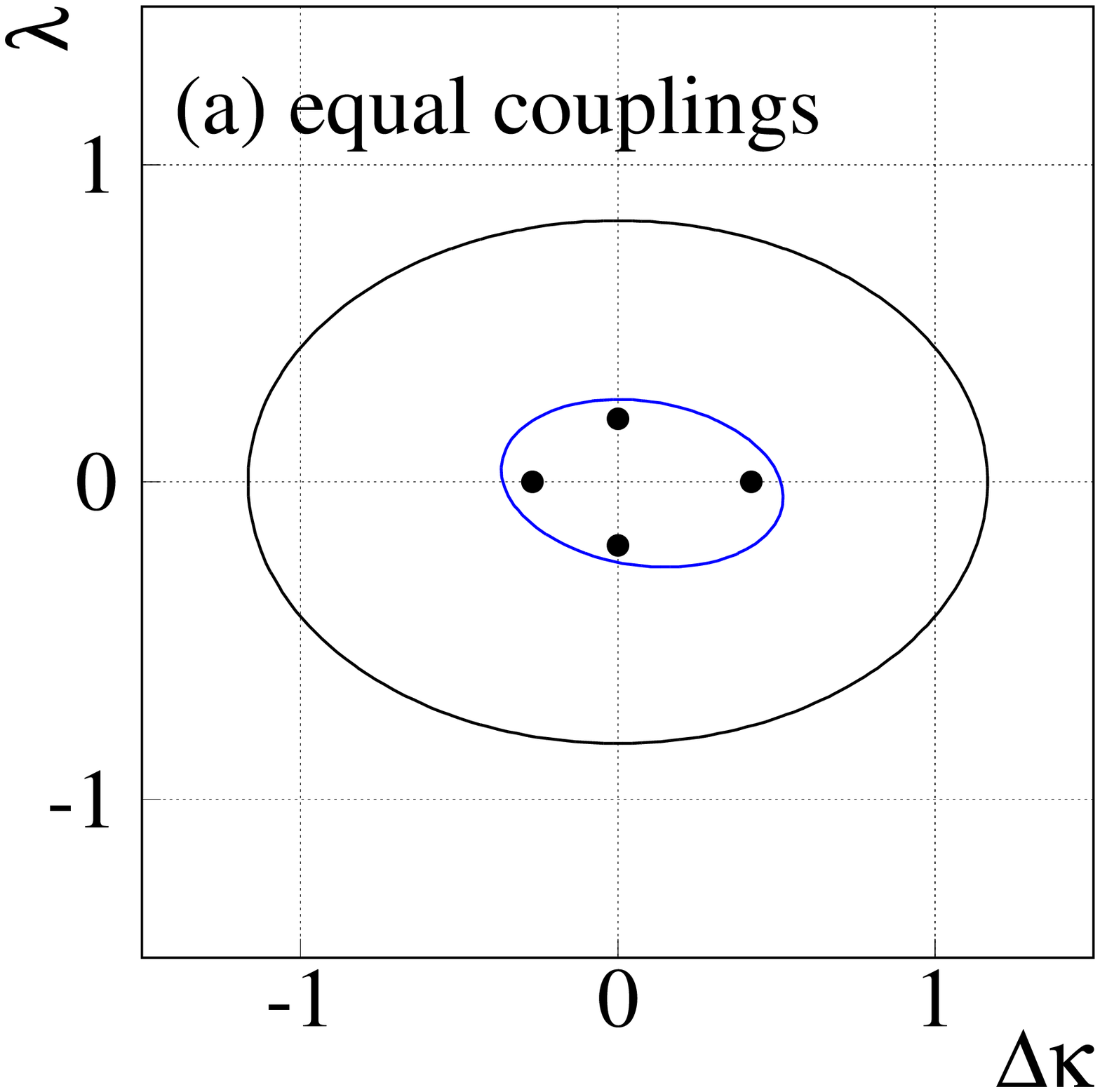}
\epsfxsize = 1.7in
\epsffile{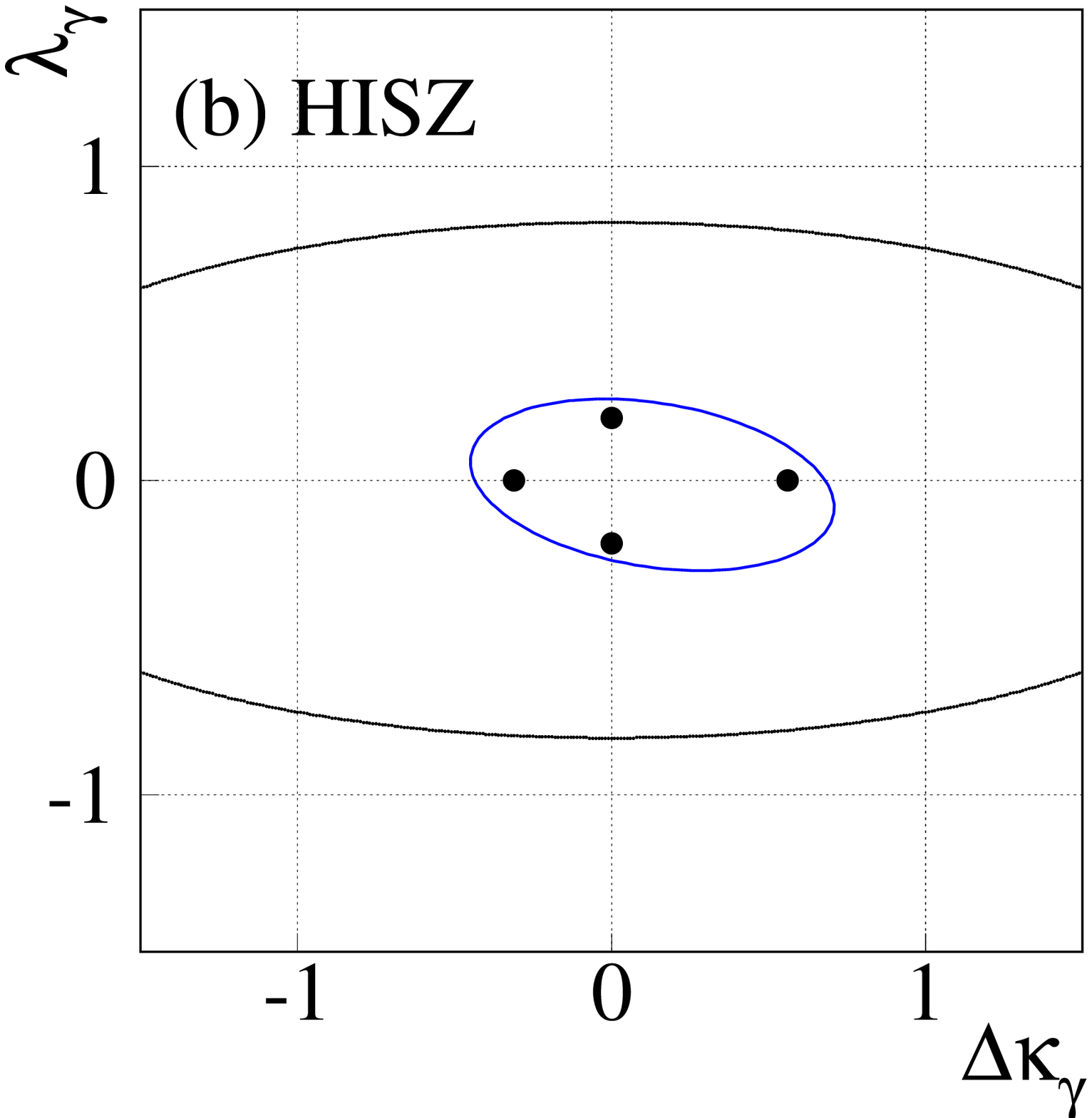}  }
\hbox{
\epsfxsize = 1.7in
\epsffile{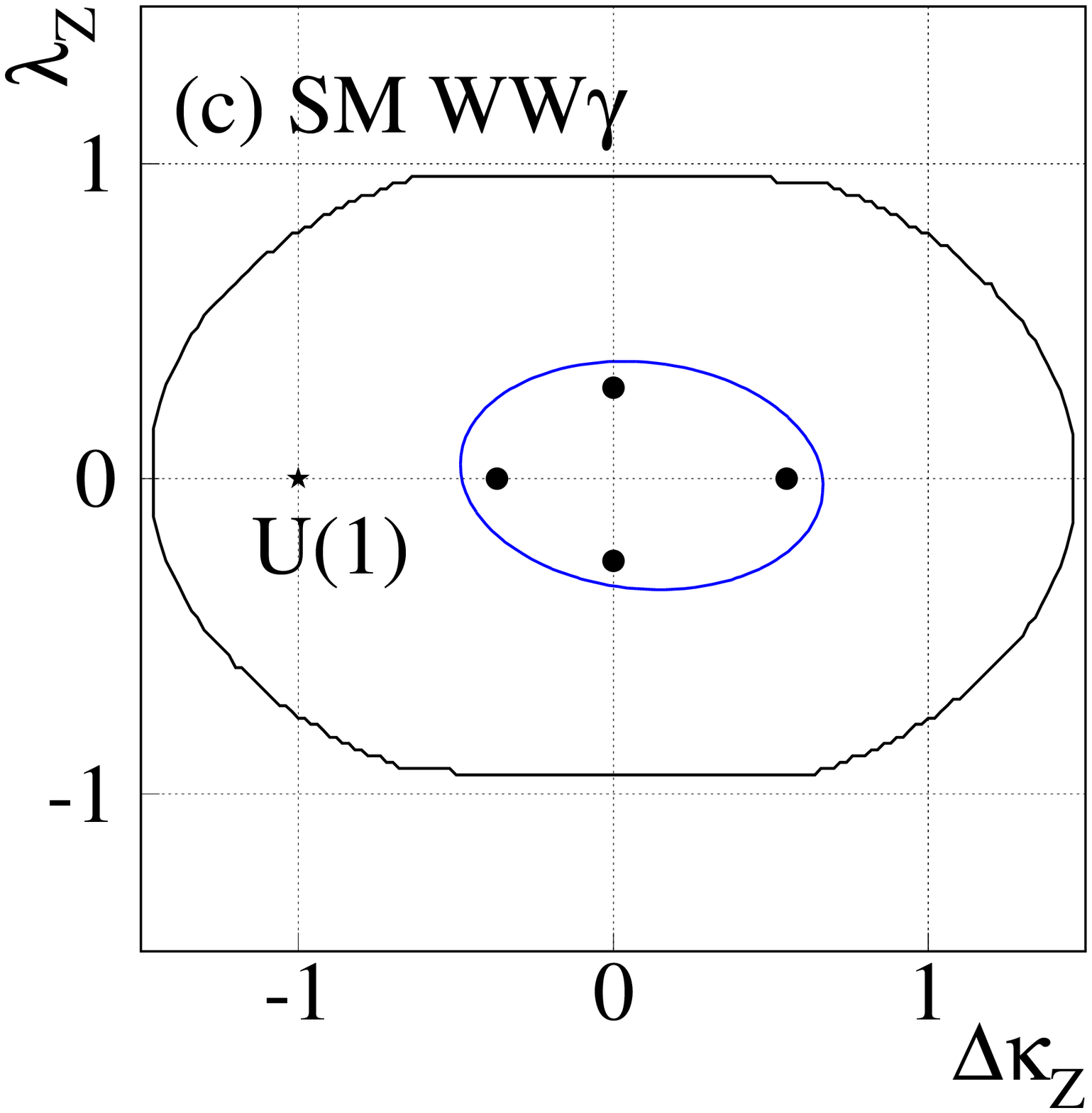}
\epsfxsize = 1.7in
\epsffile{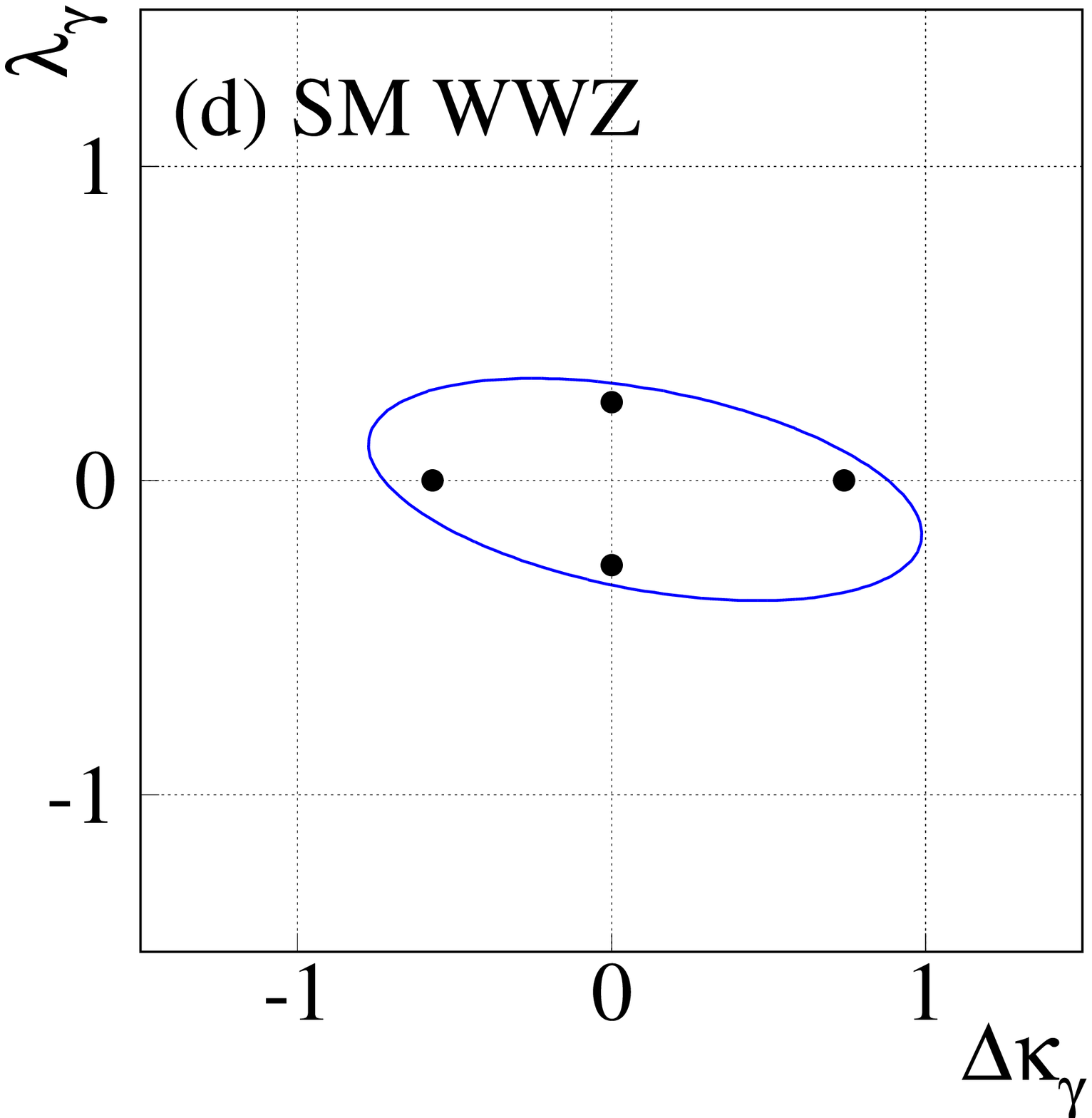} }
 \caption{Contour limits on anomalous couplings
 from a simultaneous fit to the D\O \ $W\gamma$, $WW\rightarrow$
 dilepton, $WW/WZ\rightarrow e\nu {\rm jj}$, 
 $WW/WZ\rightarrow \mu\nu {\rm jj}$, and $WZ\rightarrow$ trilepton final states
 for $\Lambda = 1.5$ TeV:
 (a) $\Delta\kappa \equiv \Delta\kappa_\gamma = \Delta\kappa_Z,
 \lambda \equiv \lambda_\gamma = \lambda_Z$; (b) HISZ relations;
 (c) SM $WW\gamma$ couplings; and (d) SM $WWZ$ couplings.
 (a), (c), and (d) assume that $\Delta g_1^Z=0$.
 The solid circles correspond to $95 \%$ C.L. one-degree of freedom exclusion
 limits. 
 The inner and outer curves are the
 $95 \%$ C.L. two-degree of freedom exclusion contour
 and the constraint from the unitarity condition, respectively.
 In (d), the unitarity contour is located outside of the
 boundary of the plot. The HISZ results include the 
 additional constraint $\alpha_{B\phi} = \alpha_{W\phi}$.}
\label{fig:combined}
\end{figure}

\begin{figure}[htb]
\hbox{
\epsfxsize = 1.8in
\epsffile{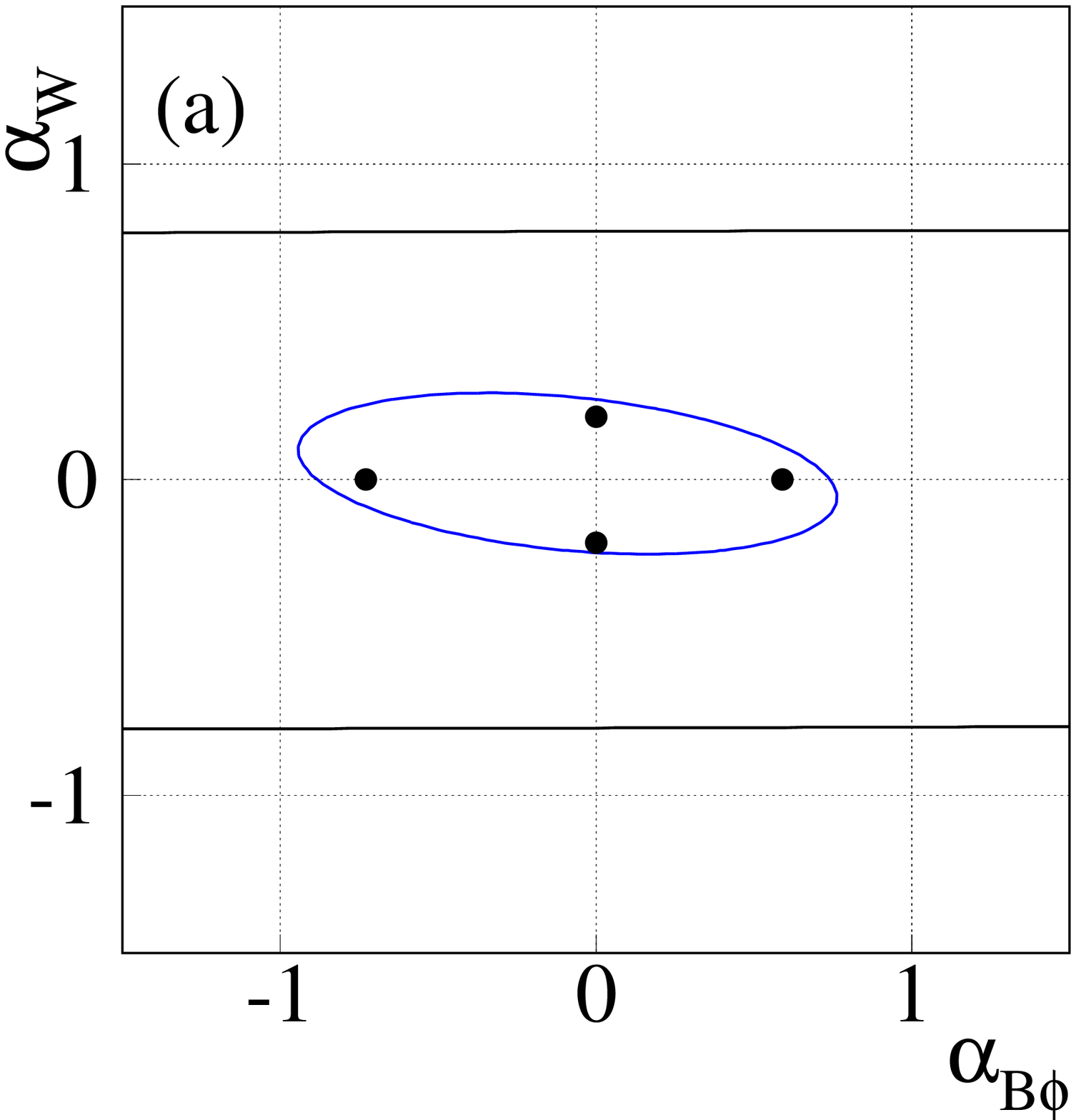}
\epsfxsize = 1.8in
\epsffile{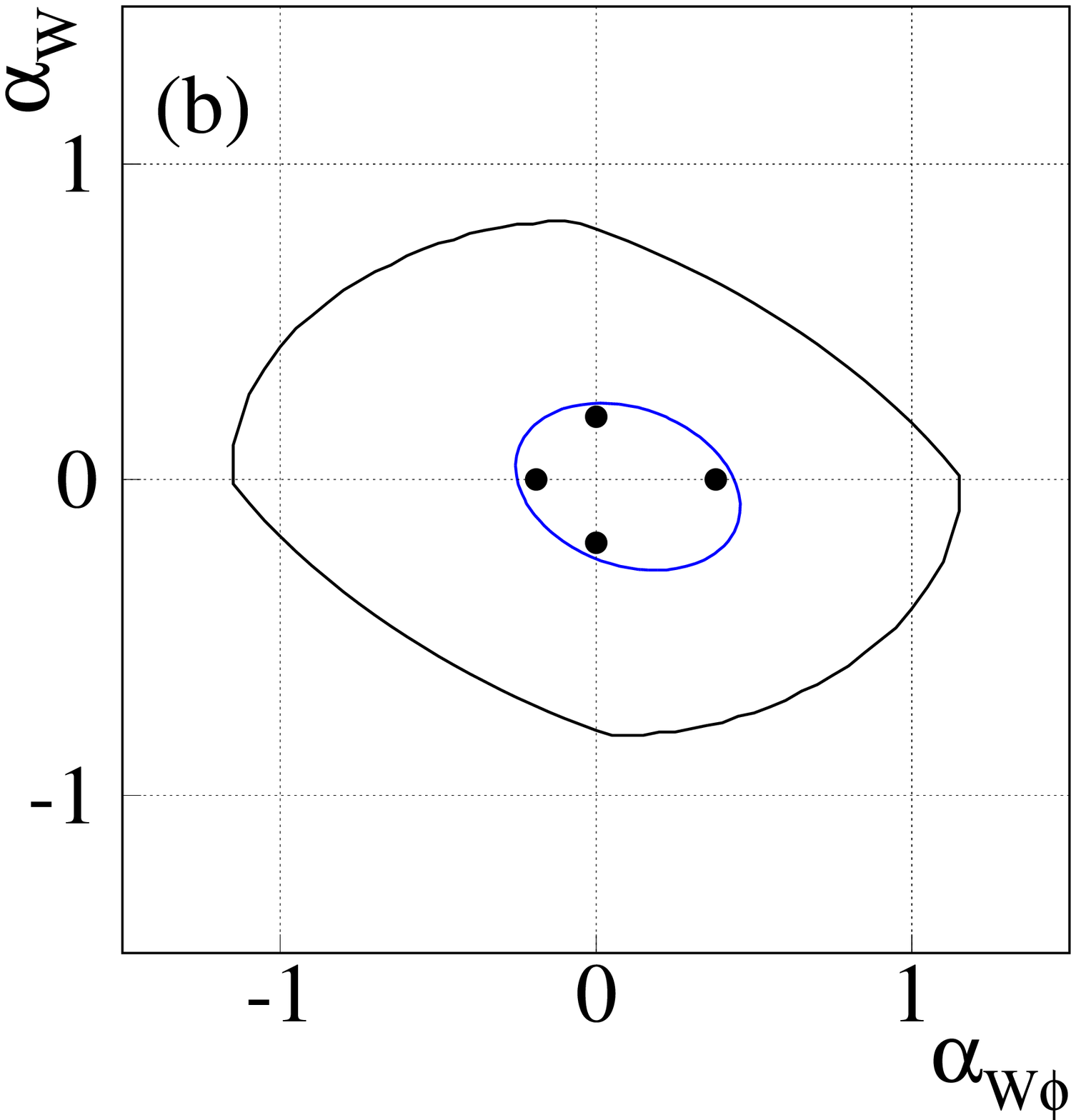}}
 \caption{Contour limits on anomalous couplings
 from a simultaneous fit to the D\O \ $W\gamma$, $WW\rightarrow$
 dilepton, $WW/WZ\rightarrow e\nu {\rm jj}$, 
 $WW/WZ\rightarrow \mu\nu {\rm jj}$, and $WZ\rightarrow$ trilepton final states
 for $\Lambda = 1.5$ TeV:
 (a) $\alpha_{W}$ vs. $\alpha_{B\phi}$ when $\alpha_{W\phi} = 0$; and
 (b) $\alpha_{W}$ vs. $\alpha_{W\phi}$ when $\alpha_{B\phi} = 0$.
 The solid circles correspond to $95 \%$ C.L. one-degree of freedom exclusion
 limits. 
 The inner and outer curves are the
 $95 \%$ C.L. two-degree of freedom exclusion contour
 and the constraint from the unitarity condition, respectively.}
 \label{fig:lep_sets}
\end{figure}

Sets of limits were produced using the range of assumptions about the
relations between the couplings as discussed in Section~\ref{sec-intro}.
Table~\ref{table-combined_limits} contains limits on $\lambda$, $\Delta
\kappa$, and where applicable on $\Delta g_1^Z$, for $\Lambda = 1.5$ and
$2.0$ TeV under each of the following assumptions: 
that the $WW\gamma$ couplings
were equal to the $WWZ$ couplings; that the $WW\gamma$ couplings were
related to the $WWZ$ couplings through the HISZ equations (with the
additional constraint $\alpha_{B\phi} = \alpha_{W\phi}$); that the
$WW\gamma$ couplings were at the SM values (producing limits on the $WWZ$
couplings); and that the $WWZ$ couplings were at the SM values (producing
limits on the $WW\gamma$ couplings).
Figure~\ref{fig:combined} shows the two-dimensional
 limit contours and one-dimensional limit points
for $\lambda$ vs. $\Delta\kappa$ for these four relationships between
the $WW\gamma$ and $WWZ$ couplings.
Table~\ref{table-combined_lims2} contains limits on
$\alpha_{B\phi}$, $\alpha_{W\phi}$, $\alpha_W$, and $\Delta g_1^Z$
for $\Lambda = 1.5$ and $2.0$ TeV.
Figure~\ref{fig:lep_sets} shows the two-dimensional limit contours and 
one-dimensional limit points
for $\alpha_{W}$ vs. $\alpha_{B\phi}$ when $\alpha_{W\phi} = 0$ and
for $\alpha_{W}$ vs. $\alpha_{W\phi}$ when $\alpha_{B\phi} = 0$.
Note that the Fig.~\ref{fig:lep_sets}(a) limits on $\alpha_{W}$ vs.
$\alpha_{B\phi}$ are equivalent to limits on $\lambda_{\gamma}$ vs.
$\Delta\kappa_{\gamma}$ because $\Delta g_1^Z$ is fixed to zero.
Also, for purposes of comparison with LEP experiments, the central values
and 68\% C.L. limits on $\lambda_{\gamma}$ and $\Delta \kappa_{\gamma}$
were calculated under the HISZ relations (without the extra constraint
$\alpha_{B \phi} = \alpha_{W \phi}$) for $\Lambda=2.0$ TeV. They were
$\lambda_{\gamma} =  0.00 ^{+0.10}_{-0.09}$
and $\Delta\kappa_{\gamma} =-0.08^{+0.34}_{-0.34}$.

\section{Conclusions}
Using $p\bar{p}$ collisions at center-of-mass energy $\sqrt{s}=1.8$ TeV 
detected with the D\O \ detector, two gauge boson pair production processes 
were studied and used to produce limits on anomalous trilinear gauge boson 
couplings.  

A search for $WZ\rightarrow e\nu ee$ and $\mu\nu ee$ 
candidates yielded one candidate event where the expected signal from SM 
$WZ$ production was $0.25\pm 0.02$ events and the expected background was 
$0.50\pm0.17$ events. The 95\% C.L. upper limit on the cross section
was $47$ pb, consistent with, but rather larger than the expected SM
 cross section. Based on the one observed event, the detection efficiency,
and the expected background, limits on anomalous $WWZ$ couplings were 
produced. The one-dimensional limits, at 95\% C.L., are 
$|\Delta g_1^Z|\le 1.63 \; (\lambda_Z=0)$ and $|\lambda|\le 1.42 \;
(\Delta g_1^Z =0)$ for $\Lambda=1.0$ TeV. 

A search for anomalous $WW/WZ\rightarrow \mu\nu {\rm jj}$ production 
was performed.  The expected background of $224\pm 56$ events was much 
larger than the 
expected SM $WW$ and $WZ$ signal of $4.5\pm 0.8$ events.  From the 
$p_T(\mu\nu)$ distribution of the 224 observed
events, which had no significant deviation
from the expected background plus SM signal,  limits on 
anomalous $WW\gamma$ and $WWZ$ couplings were produced.  
Under the assumption that
the $WW\gamma$ couplings equal the $WWZ$ couplings, the one-dimensional
95\% C.L. limits were $-0.43 \le \lambda \le 0.44 \; (\Delta \kappa = 0)$ 
and $-0.60 \le \Delta \kappa \le 0.74 \; (\lambda=0)$ for $\Lambda=2.0$
TeV. Under the assumption that the $WW\gamma$ couplings are related to the 
$WWZ$ couplings via the HISZ equations, the one-dimensional
95\% C.L. limits were $-0.42 \le \lambda \le 0.44 \; (\Delta \kappa = 0)$
and $-0.71 \le \Delta \kappa \le 0.96 \; (\lambda=0)$ for $\Lambda=2.0$
TeV. 

The results of the two searches described in this paper have been combined 
with those from our previous publications to produce our most restrictive
limits on anomalous $WW\gamma$ and $WWZ$ couplings. Under the assumption that
the $WW\gamma$ couplings equal the $WWZ$ couplings, the one-dimensional
95\% C.L. limits were $-0.18 \le \lambda \le 0.19 \; (\Delta \kappa = 0)$
and $-0.25 \le \Delta \kappa \le 0.39 \; (\lambda=0)$ for $\Lambda=2.0$
TeV. Under the assumption that the $WW\gamma$ couplings are related to the
$WWZ$ couplings via the HISZ equations, the one-dimensional
95\% C.L. limits were $-0.18 \le \lambda \le 0.19 \; (\Delta \kappa = 0)$
and $-0.29 \le \Delta \kappa \le 0.53 \; (\lambda=0)$ for $\Lambda=2.0$
TeV.  Limits on $\Delta \kappa$, $\lambda$, and $\Delta g_1^Z$ were 
determined for the $WW\gamma$ couplings assuming the $WWZ$ couplings
are at the SM value, and for the $WWZ$ couplings assuming that the $WW\gamma$
couplings are  at the SM value.  Finally, limits on the $\alpha_{B\phi}$,
$\alpha_{W\phi}$, and $\alpha_W$ anomalous couplings were produced.  

%
We thank the Fermilab and collaborating institution staffs for
contributions to this work and acknowledge support from the 
Department of Energy and National Science Foundation (USA),  
Commissariat  \` a L'Energie Atomique (France), 
Ministry for Science and Technology and Ministry for Atomic 
   Energy (Russia),
CAPES and CNPq (Brazil),
Departments of Atomic Energy and Science and Education (India),
Colciencias (Colombia),
CONACyT (Mexico),
Ministry of Education and KOSEF (Korea),
and CONICET and UBACyT (Argentina).

\section{Appendix: Parameters of the $WZ$ Candidate Event}
Given the expected signal to background ratio of
approximately one to two in the channel $WZ\rightarrow  e\nu ee$, there is 
no certainty that the candidate event is actually due to $WZ$ production. 
But due to the event's striking signature it is described in detail in this 
appendix.

The candidate event contains three high-$E_T$ electron candidates and large
missing transverse energy (46.2 GeV).  The event contains no other high-$p_T$ 
objects (jets or muons). The properties of the candidate electrons are
summarized in Table~\ref{tabl.candelec}. The missing transverse energy and 
the various mass combinations of the electrons with the missing transverse 
energy are listed in Table~\ref{tabl.candmass}. The invariant  mass of 
electron  candidates 1 and 3 is $93.6$ GeV/$c^2$,  and the transverse 
mass formed 
using electron candidate 2 and the missing transverse energy is 
74.7 GeV/$c^2$.

\twocolumn

\begin{table}
\begin{center}
\begin{tabular}{|c|c|c|}
Electron Type & Efficiency (CC) \%  & Efficiency (EC) \%  \\ \hline \hline
Loose         & $88.6 \pm 0.3$ & $88.4 \pm 0.5$ \\ \hline
Tight         & $73.4 \pm 0.5$ & $67.2 \pm 0.3$ \\
\end{tabular}
\end{center}
\caption{Measured efficiencies for electron identification in the CC and two
EC's.  See text for definitions of tight and loose.}
\label{table.eideff}
\end{table}

\begin{table}
\begin{center}
\begin{tabular}{|c||c|c||c|c|}
Electron & \multicolumn{2}{|c||}{CC} & \multicolumn{2}{|c|}{EC} \\ \cline{2-5}
Type     &  $a_{0}\times 10^{3}$   & $a_{1}\times 10^{5}$  &   $a_{0}\times
                         10^{3} $  & $a_{1}\times 10^{5}$ \\ \hline
\hline
Loose
& $0.08 \pm 0.29$ & $2.06 \pm 0.70$ & $1.3 \pm 1.0$ & $6.31 \pm 0.27$ \\ \hline
Tight
& $-0.17 \pm 0.20$ & $1.43 \pm 0.51$ & $0.53 \pm 0.86$ & $5.1 \pm 2.3$ \\
\end{tabular}
\caption{Jet misidentification probabilities for tight and
loose electrons. The probability is a linear function of $E_T({\rm GeV})$, 
$a_0+a_1 \times E_T({\rm GeV})$. 
Uncertainties given in this table are statistical only.
A systematic uncertainty of 25\% was assigned to each fake
probability. }
\label{table-pje}
\end{center}
\end{table}

\begin{table}
\begin{center}
\begin{tabular}{|c|c c |c|}
              & $e\nu ee$ & $\mu\nu ee$& Total \\ \hline
${\cal L}$ & \multicolumn{2}{c|}{$92.3\pm 5.0$~pb$^{-1}$}&\\
$\epsilon$ & $0.169\pm 0.014$& $0.115\pm 0.014$&\\
$Br$ & \multicolumn{2}{c|} {$0.36\%\pm 0.01\%$}&\\
$N_{{\rm obs}}$ &  $1$        &   $0$&  $1$      \\
$N_{{\rm bkg}}$ & $0.38\pm 0.14$ & $0.12\pm 0.04$ & $0.50\pm0.17$\\
$N_{{\rm SM}}$&$0.15\pm0.01$&$0.10\pm0.01$&$0.25\pm0.02$\\
\end{tabular}
\end{center}
\caption{Summary of the $WZ\to l\nu ll$ results. ${\cal L}$ is the integrated
luminosity, $\epsilon$ is the overall
detection efficiency, $Br$ is the branching ratio, $N_{{\rm obs}}$ 
is the number of
events observed, $N_{{\rm bkg}}$ is the number of background events, 
and $N_{{\rm SM}}$ is
the predicted number of SM events.}
\label{results}
\end{table}

\begin{table}[htb]
\begin{tabular}{l|c}
\multicolumn{1}{l|}{Sample} &
\multicolumn{1}{c}{Number of Events} \\ \hline \hline
QCD multi-jet background & 105$\pm$19 (stat)\\ \hline
W + $>$ 2 jets background & 117$\pm$24 (stat)\\ \hline
$t\bar{t}$ background & 2.7$\pm$1.2 (stat)\\ \hline
Total background & 224$\pm$31 (stat) $\pm 46$ (syst)\\ \hline
SM prediction & 4.5$\pm$0.8 (stat + syst) \\ \hline
Observed data sample &       \\
(luminosity = $80.7$ pb$^{-1}$) & 224 \\
\end{tabular}
\caption{Comparison of signal (data) and backgrounds for the mode
$WW/WZ \to \mu\nu {\rm jj}$. The data sample is consistent with the SM
prediction and estimated backgrounds showing no evidence for anomalous
gauge couplings.}
\label{tab-evsum}
\end{table}

\begin{table}[htb]
\begin{tabular}{l|c|c}
\multicolumn{1}{l|}{Coupling} &
\multicolumn{1}{c|}{$\Lambda=1.5$ TeV} &
\multicolumn{1}{c}{$\Lambda=2.0$ TeV} \\ \hline \hline
$\lambda_\gamma=\lambda_Z$ & $-0.45$, 0.46 & $-0.43$, 0.44 \\ \hline
$\Delta\kappa_\gamma=\Delta\kappa_Z$ & $-0.62$, 0.78 & $-0.60$, 0.74 \\ \hline
$\lambda_\gamma=\lambda_Z$(HISZ) & $-0.44$, 0.46 & $-0.42$, 0.44 \\ \hline
$\Delta\kappa_\gamma$(HISZ) & $-0.75$, 0.99 & $-0.71$, 0.96 \\
\end{tabular}
\caption{Axis limits (one-dimensional) 
at the 95\% C.L. with two assumptions for the 
relation between the 
$WW\gamma$ and $WWZ$ couplings $(WW\gamma=WWZ$ and HISZ) and for two
different values of $\Lambda$ in the mode $WW/WZ \to \mu\nu {\rm jj}$.}
\label{tab-limits}
\end{table}
  
\begin{table*}[htb]
\begin{center}
\begin{tabular}{|c|c|c|}
Coupling &$\Lambda=1.5$ TeV&$\Lambda=2.0$ TeV\\ \hline
$\lambda_{\gamma} = \lambda_Z$ ($\Delta\kappa_{\gamma}=\Delta\kappa_Z=0$)&
 $-0.20,~0.20$& $-0.18,~0.19$\\
$\Delta\kappa_{\gamma} = \Delta\kappa_Z$ ($\lambda_{\gamma}=\lambda_Z=0$)&
 $-0.27,~0.42$& $-0.25,~0.39$\\ \hline
$\lambda_{\gamma}$(HISZ) ($\Delta\kappa_{\gamma}=0$)&
 $-0.20,~0.20$& $-0.18,~0.19$\\
$\Delta\kappa_{\gamma}$(HISZ) ($\lambda_{\gamma}=0$)&
 $-0.31,~0.56$& $-0.29,~0.53$\\ \hline
$\lambda_Z$(SM $WW\gamma$) ($\Delta\kappa_Z=\Delta g^Z_1=0$)&
 $-0.26,~0.29$& $-0.24,~0.27$\\
$\Delta\kappa_Z$(SM $WW\gamma$) ($\lambda_Z=\Delta g^Z_1=0$)&
 $-0.37,~0.55$& $-0.34,~0.51$\\
$\Delta g^Z_1$(SM $WW\gamma$) ($\lambda_Z=\Delta\kappa_Z=0$)&
 $-0.39,~0.62$& $-0.37,~0.57$\\ \hline
$\lambda_{\gamma}$(SM $WWZ$) ($\Delta\kappa_{\gamma}=0$)&
 $-0.27,~0.25$& $-0.25,~0.24$\\
$\Delta\kappa_{\gamma}$(SM $WWZ$) ($\lambda_{\gamma}=0$)&
 $-0.57,~0.74$& $-0.54,~0.69$\\
\end{tabular}
\end{center}
\caption{One-dimensional 
limits at 95\% C.L. from a simultaneous fit to the D\O \ $W\gamma$,
$WW\rightarrow$ dilepton, $WW/WZ\rightarrow e\nu {\rm jj}$,
$WW/WZ\rightarrow \mu\nu {\rm jj}$, and $WZ\rightarrow$ trilepton data samples.
The HISZ results included the
 additional constraint $\alpha_{B\phi} = \alpha_{W\phi}$.}
\label{table-combined_limits}
\end{table*}
                         
\begin{table}[htb]
\begin{tabular}{|c|c|c|}
Coupling &$\Lambda=1.5$ TeV&$\Lambda=2.0$ TeV \\ \hline
$\alpha_{B\phi}$ ($\alpha_{W\phi}=\alpha_W=0$)&
 $-0.73,~0.59$& $-0.67,~0.56$ \\ \hline
$\alpha_{W\phi}$ ($\alpha_{B\phi}=\alpha_W=0$)&
 $-0.19,~0.38$& $-0.18,~0.36$ \\ \hline
$\alpha_W$ ($\alpha_{B\phi}=\alpha_{W\phi}=0$)&
 $-0.20,~0.20$& $-0.18,~0.19$ \\ \hline
$\Delta g^Z_1$ ($\alpha_{B\phi}=\alpha_W=0$)&
 $-0.25,~0.49$& $-0.23,~0.47$ \\
\end{tabular}
\caption{One-dimensional 
limits at 95\% C.L. on $\alpha$ parameters from a simultaneous fit
to the D\O \ $W\gamma$, $WW\rightarrow$ dilepton, $WW/WZ\rightarrow e\nu 
{\rm jj}$,
$WW/WZ\rightarrow \mu\nu {\rm jj}$, 
and $WZ\rightarrow$ trilepton data samples.}
\label{table-combined_lims2}
\end{table}

\begin{table}[h]
\begin{center}
\begin{tabular}{|c| c c c|}
             & $e_1$ & $e_2$ & $e_3$\\
\hline
$E_T$ (GeV)  & 54.5  & 50.9  & 37.7 \\
$\eta$       & 0.11  & $-0.62$ & 1.37 \\
$\phi$       & 5.94  & 3.04  & 4.14 \\
\end{tabular}
\caption{Kinematic properties of the $WZ\rightarrow  e\nu ee$ candidate
event (Run 89912, Event 23020). }
\label{tabl.candelec}
\end{center}
\end{table}

\begin{table}[h]
\begin{center}
\begin{tabular}{l l}
\multicolumn{2}{c}{Mass Combination Information} \\ \hline
 $M_{e_1,e_2} = 111.8$ GeV/$c^2$ & $M_{e_1,e_2,e_3}   =  171.7$ GeV/$c^2$ \\
 $M_{e_1,e_3} =  93.6$ GeV/$c^2$ & $M_{e_2,e_3}  =  112.4$ GeV/$c^2$ \\
 $\hbox{$\rlap{\kern0.25em/}E_T$} =  46.2$ GeV
                           &
             $\phi(\hbox{$\rlap{\kern0.25em/}E_T$}) =     1.29$  \\
 $M_T(e_i,\hbox{$\rlap{\kern0.25em/}E_T$}) = 73.0,  74.7,  82.6$   GeV/$c^2$
                           &  for $e_1$, $e_2$, $ e_3$ respectively \\
 $p_T(e_1,e_3) =  58.8$ GeV/$c$    & $\phi(e_1,e_3) = -1.02$ \\
 $p_T(e_2,\hbox{$\rlap{\kern0.25em/}E_T$}) =  63.0$ GeV/$c$
                           &
             $\phi(e_2,\hbox{$\rlap{\kern0.25em/}E_T$})=  2.22$ \\
\end{tabular}
\caption{Mass combination information for $e\nu ee$ candidate event.
$M_{e_i,e_j}$ is the invariant mass of electron $i$ and electron $j$. $M_{e_1,
e_2,e_3}$ is the three-body mass of electron 1, electron 2, and electron 3.
$M_T$ is the transverse mass and  $p_T$ is the transverse momentum.}
\label{tabl.candmass}
\end{center}
\end{table}

\clearpage

\begin{references}
\bibitem{newref} Limits on anomalous couplings, presented in this paper, 
 are given at the low energy limits of the couplings.
\bibitem{pseudo} Pseudorapidity, $\eta$, is defined as $-\ln{\tan{\theta /2}}$
where $\theta$ is the polar angle. In the definition of
  the pseudorapidity of the calorimeter detector boundaries, the origin was
  the center of the calorimeter.  In the definition of the pseudorapidity
  of an object such as an electron or muon, the origin was the $p\bar{p}$
  collision point.
\end{references}
\end{document}